\documentclass[12pt]{article}
\usepackage{amsmath,amssymb,amsthm,color}
\usepackage{graphicx}
\usepackage{cite}
\usepackage{epsfig}
\usepackage{lineno}
\renewcommand{\baselinestretch}{1.66}

\begin{document}
\title{Coulomb excitations and decays in graphene-related systems \\}
\author{Chiun-Yan Lin$^{a}$, Jhao-Ying Wu$^{b}$, Chih-Wei Chiu$\dag$$^{c}$, and Ming-Fa
Lin$\ddag$$^{a,d,e}$\\
\small  $^a$Department of Physics, National Cheng Kung University, Tainan, Taiwan\\
\small  $^b$Center of General Studies, National Kaohsiung University of Science and Technology,
Kaohsiung, Taiwan\\
\small  $^c$Department of Physics, National Kaohsiung Normal University, Kaohsiung, Taiwan\\
\small  $^d$Hierarchical Green-Energy Materials Research Center, National Cheng Kung University, Tainan,
Taiwan\\
\small  $^e$Quantum topology center, National Cheng Kung University, Tainan, Taiwan\\
}
\renewcommand{\baselinestretch}{1.66}
\maketitle

\renewcommand{\baselinestretch}{1.66}

\begin{abstract}
The layered graphene systems exhibit the rich and unique excitation spectra arising from the
electron-electron Coulomb interactions. The generalized tight-binding model is developed to cover the
planar/buckled/cylindrical structures, specific lattice symmetries, different layer numbers, distinct
configurations, one-three dimensions, complicated intralayer and interlayer hopping integrals, electric
field, magnetic quantization; any temperatures and dopings simultaneously. Furthermore, we modify the
random-phase approximation to agree with the layer-dependent Coulomb potentials with the Dyson equation,
so that these two methods can match with other under various external fields. The electron-hole
excitations and plasmon modes are greatly diversified by the above-mentioned critical factors; that is,
there exist the diverse (momentum. frequency)-related phase diagrams. They provide very effective
deexcitation scatterings and thus dominate the Coulomb decay rates. Graphene, silicene and germanene
might quite differ from one another in Coulomb excitations and decays because of the strength of
spin-orbital coupling. Part of theoretical predictions have confirmed the experimental measurements, and
most of them require the further examinations. Comparisons with the other models are also made in
detail.
\end{abstract}

\par\noindent  $\dag$ Corresponding author.
{~ Tel:~ +886-7-717-2930.}\\~{{\it E-mail addresses}:
giorgio@mail.nknu.edu.tw (C. W. Chiu),
\par\noindent  $\ddag$ Corresponding author.
{~ Tel:~ +886-6-275-7575.}\\~{{\it E-mail addresses}:
mflin@mail.ncku.edu.tw (M. F. Lin)}

\pagebreak
\renewcommand{\baselinestretch}{2}
\newpage

\vskip 0.6 truecm



\setcounter{page}{1}
\pagenumbering{arabic}

\section{Introduction}

The graphene-related systems have attracted a lot of theoretical and experimental researches, mainly
owing to the unique hexagonal symmetry, unusual layered structures/nanoscaled thickness, and various
stacking configurations.\cite{ACSNano4;6337,AdvMater22;3723,ElecComm11;889,ACSNano4;1790,
PRL97;187401,Science306;666,RevModPhys81;109,ChemRev110;132,Electroanalysis22;1027,
ACSCatal2;781,AccChemRes46;2329,RPP76;056503} Such systems are very suitable for exploring the basic
science and the high-potential applications.\cite{ACSNano4;6337,AdvMater22;3723,ElecComm11;889,
ACSNano4;1790,AccChemRes46;2329,Electroanalysis22;1027,PRL97;187401,Science306;666,RevModPhys81;109,
NatNanotechnol7;330,NatPhotonics7;394,ACSNano8;1086,ACSPhotonics2;151,ACSPhotonics5;1612} Electronic
excitations/deexcitations arising from the electron-electron Coulomb interactions are one of essential
physical properties, being closely related to the geometric and band structures. They are determined by
the intrinsic many-particle properties and play a critical role in all the condensed-matter systems with
the different dimensions.  A new theoretical framework is developed for the layered systems; a modified
random-phase approximation (RPA) is conducted on the 2D materials. In this book, a systematic and
thorough investigation clearly indicate the diverse Coulomb excitation/decay phenomena, in which it
covers many ${sp^2}$-bonding carbon-created materials, e.g., monolayer graphene, double-layer graphene,
AA-, AB- ABC- $\&$ AAB-stacked graphenes, sliding bilayer graphene, simple graphite, Bernal graphite,
rhombohedral graphite, metallic/semiconducting carbon nanotubes, and monolayer silicene/germanene. The
composite effects due to lattice symmetry, layer number, dimensionality, stacking configuration,
temperature, doping, electric field, magnetic field will be discussed in detail. Specially, the
magneto-electronic excitations, which are associated with the magnetic quantization, require the
combination of the generalized tight-binding mode and the modified RPA. The theoretical predictions are
fully compared with the experimental measurements from the electron energy loss spectroscopy (EELS), and
part of them are consistent with the latter. The measured EELS spectra have confirmed the diverse
collective excitations (plasmon modes) of the free carriers, and the $\pi$ $\&$ ${\pi\,+\sigma}$ valence
electrons at different frequency ranges (the detailed discussions in Chap. 2.5).

The dielectric function ($\epsilon$) and the dimensionless energy loss (Im${[-1/\epsilon\,]}$) function,
which, respectively, represent the bare and screened response ability of charged carriers under the
Coulomb field perturbation, are critical in fully understanding the intrinsic excitation/decay
properties/of condensed-matter materials. In general, the imaginary part of $\epsilon$ and the prominent
peaks in Im${[-1/\epsilon\,]}$, respectively, correspond to the single-particle and collective
excitations (electron-hole excitations and plasmon modes). $\epsilon$ is strictly defined for bulk
graphites,\cite{PRB34;2} monolayer graphene,\cite{PRB34;2} and cylindrical carbon
nanotubes,\cite{JPSJ66;757} since such systems possess a good translational symmetry, respectively, in
the 3D, 2D and 1D spaces.
That is to say, the bare Coulomb electron-electron interactions mainly determined by the transferred
momentum (${\bf q}$) exhibit well-behaved/dimension-determined forms. On the other hand, the longitudinal
dielectric function is expressed as a tensor form for the layered graphenes. The layer-projection method,
which is closely related to the Coulomb scattering of the initial and final electronic states, will be
introduced to deal with the electric polarizations. As a result, $\epsilon$ is a layer-dependent tensor
function with the double indices ($\epsilon_{ll^\prime}$; $l$ the layer index), in which the
band-structures effects on the excitation spectra and Coulomb matrix elements are fully taken into
account in the theoretical model. For example, a $N$-layer graphene possesses $N$ pairs of valence and
conduction bands due to the carbon ${2p_z}$ orbitals, and the sublattice-decomposed wave functions are
utilized to accurately evaluate the layer-dependent ${P_{ll^\prime}}$ or ${\epsilon_{ll^\prime}}$.
Under the Born approximation, the effective energy loss function is characterized by the inelastic
scattering probability of the incident electron beam. It has been very successfully in understanding the
diverse electronic excitation spectra in layered systems, e.g.,
AA-,\cite{JPSJ81;124713,PRB86;125434,AOP339;298,PRB74;085406}
AB-\cite{PRB74;085406,SciRep3;1368,PLA352;446} and ABC-stacked graphenes.\cite{PRB98;041408}
The magnetoplasmon is discussed in this book by the development of the modified RPA and the generalized
Peierls tight-binding model.\cite{IOPBook;978,IOPChen,PCCP17;26008,CRCPress;9781138571068}
It is very suitable for studying the inter-landau-level (inter-LL), single-particle excitations and
magneto-plasmon modes.\cite{ACSNano5;1026,PRB89;165407,PRB94;205427} Moreover, electronic excitation
spectra are also quite efficient decay channels by the inelastic Coulomb scatterings, being the strong
effects on the energy widths of the quasiparticle states (the excited electrons or holes). The Coulomb
decay rates are evaluated from the self-energy method under the detailed derivations. They will show
certain important differences among monolayer graphene, silicene and germanene with the electron/hole
doping. As to the experimental side, EELS and inelastic X-ray scatterings (IXS) are two very powerful
techniques in providing Coulomb excitation phenomena. Their recent developments \cite{Egerton,Winfried}
and measurements on the screened response functions of graphene-related
materials\cite{OptCom1;119,ZPhys243;229,PRL100;196803,PRB88;075433} are explored in detail. Also, the
quasi-particle energy spectra under the measurements of the angle-resolved photo-emission spectroscopy
(ARPES) are useful in the determinging the Coulomb decay
rates.\cite{PRL102;107007,NatComm5;3257,NatPhys3;36}

Monolayer graphene, as displayed in Fig. 1-1(a), has a planar honeycomb lattice composed of two
equivalent sublattices, so that it exhibits the linear and isotropic Dirac-cone bans structure near the
Fermi level ($E_F$).\cite{RevModPhys81;109} A pristine system is only a zero-gap semiconductor, since
density of states (DOS) is vanishing at $E_F$. The low-lying DOS is a linear V-shape form centered at
${E_F=0}$.\cite{RevModPhys81;109} It leads to the specific interband e-h excitations and the absence of
intraband ones.\cite{RevModPhys81;109} Temperature could induce the thermal excitations between the
gapless valence and conduction states; that is, it generates conduction electrons and valence holes. The
free carrier density per area is identified to reveal a simple $T^2$-dependence, and their magnitude is
estimated to ${\sim\,10^{11}}$ e/cm$^2$ at room temperature. The intraband single-particle excitations
are induced/enhanced by temperature, while the great reduce is observed in the interband ones. Most
important, temperature can create intraband plasmon modes at sufficiently high T, in which they belong to
2D acoustic collective excitations defined by a specific $\sqrt q$-dependence at long wavelength limit.
By means of the alkali-adatom absorptions or the applications of gate
voltages,\cite{AdvPhys51;1,PRB81;115428,Carbon57;507,Nanotechnol22;425701} there exists a very high
free-carrier density (${\sim\,10^{13}}$ e/cm$^2$) in an extrinsic graphene system.
The intraband $\&$ interband excitations and the intraband plasmons are expected to be very pronounced in
the bare and screened excitation spectra, respectively.
This has been clearly confirmed from the EELS measurements.\cite{PRB88;075433}
Monolayer band structure is also reliable when the interlayer distance is very large (e.g., more than
double that of graphite).\cite{PRB50;17744}
The effects, which purely arise from the interlayer Coulomb interactions, could be investigated for a
double-layer graphene system.
The acoustic and optical plasmon modes are clearly revealed in the EELS spectra, and they, respectively,
correspond to the collective oscillations of two-layer charge carriers in phase and out of
phase.\cite{PRB80;241402,PRL81;4216} The experimental measurements are required to verify these two
modes.

The layered graphenes present the various stacking configurations, such as, AAA, ABA, ABC and AAB ones.
The symmetries of geometric structures have strong effects on band structures and thus greatly diversify
Coulomb excitations/decays. The AA stacking means that carbon atoms in different layers possess the same
${(x,y)}$ projections, as shown in Fig. 1-1(b). From the first-principles evaluations using
VASP,\cite{Carbon32;289} the ground state energy of the AA-stacked graphene is predicted to be highest
among all the stacking configurations. This clearly indicates that it is relatively difficult to
synthesize the AA stacking in the experimental laboratory.\cite{PRL102;015501} According to both VASP and
tight-binding model calculations,\cite{PCCP17;26008,CRCPress;9781138556522} there exist $N$ pairs of
Dirac-cone structures in the N-layer AA stacking; furthermore, such linear energy bands are intersecting
at the K/K$^\prime$ valley even in the presence of the vertical and non-vertical interlayer hopping
integrals. For example, two pairs of valence Dirac cones in bilayer system are identified from the
measurements of angle-resolved  photoemission spectroscopy (ARPES) on the quasi-particle energy
spectrum.\cite{NatMater23;887,NanoLett17;1564} The valence and conduction bands strongly overlap one
another so that a lot of free electrons and holes are created by the interlayer atomic interactions. This
semi-metallic property is in sharp contrast with the semiconducting behavior of monolayer graphene. Such
carriers further induce the collective excitations in which the number of plasmon modes is identical to
that of layer.\cite{JPSJ81;124713} Also, the single-particle excitation channels are complicated/enriched
by more energy bands. From the theoretical point of view, the interlayer hopping integrals and the
interlayer Coulomb interactions need to be simultaneously included in the analytic formulas; that is,
the band-structure effects are thoroughly covered in the electronic excitations/deexcitations.
The main features of plasmon modes and electron-hole (e-h) dampings, the sensitive dependences on the
transferred momentum and energy [(${\bf q\,,\omega}$)], are worthy of a systematic investigation. In
addition, the experimental measurements are absent up to now.

The AB-stacked graphenes frequently appear in the experimental observations using the various methods,
e.g., the successful syntheses by the mechanical exfoliation,\cite{Science306;666,SurSci610;53} the
chemical vapor deposition (CVD),\cite{AdvFun19;2782,ACSNano3;2653} and the electrostatic manipulation of
scanning tunneling microscopy (STM).\cite{Carbon50;4633,PRB86;085428} This shows that their ground state
energies are much lower than those of the AA stackings.\cite{SurSci610;53} The AB stacking is the natural
periodical sequence of Bernal graphite.\cite{PRSLSA106;749,PRSLSA181;101} The neighboring layers are
attracted together by the weak but significant van der Waals interactions, so that the few-layer AB
stacking could be obtained from the natural graphite under the mechanical
action.\cite{Science306;666,SurSci610;53} They only possess half of carbon atoms in the same ${(x,y)}$
projections (Fig. 1-1(c)), or they are characterized by a relative shift of the C-C bond length ($b$)
along the armchair direction about the initial AA stacking. There are more complex interlayer hopping
integrals and extra site energies due to the different chemical environments experienced by the A$^l$ and
B$^l$ sublattices ($l$ layer index), compared with the AA stacking. For AB bilayer stacking, the
Dirac-cone energy bands  become two pairs of parabolic valence and conduction bands, in which the latter
are also initiated from the K/K$^\prime$ valley (the corners of the hexagonal first Brillouin zone).
Furthermore, the trilayer system presents an extra  Dirac cone with a slight distortion. Such
band-structure characteristics of few-layer AB stackings have been confirmed by the ARPES
measurements\cite{PRL98;206802,PRB88;075406,PRB88;155439,NPGAsiaMat10;466}. In short, a even-$N$ (an
odd-$N$) AB stacking exhibits $N$ pairs of parabolic bands (accompanied with a weakly separated
Dirac-cone structure). In general, a small overlap of valence and conduction bands is revealed in the
first pair of few-layer systems.\cite{PRL98;206802} Apparently, the 2D free electron/hole density purely
arising from the interlayer interactions is very low, leading to the absence of collective excitations in
pristine AB-stacked systems.\cite{arXiv:1803;10715,PRB87;235418}  However, extrinsic few-layer AB
stackings are expected to present the unusual Coulomb excitation behaviors (Chap. 5).

The ABC stacking, corresponding to Fig. 1-1(d)), is predicted to have the lower ground state energy than
that of the ABA stacking.\cite{CRCPress;9781138556522} The ABC-stacked few-layer graphenes are easily
synthesized in the experimental growth.\cite{SurSci610;53,ACSNano3;2653,AdvFun19;2782}  However, its 3D
counterpart, rhombohedral graphite, is only presented with a low-concentration arrangement in natural
graphite; that is, most of 3D graphite belongs to the ABA (Bernal)
stacking.\cite{PRSLSA106;749,PRSLSA181;101} The unique geometric symmetry induces the rich hopping
integrals and the unusual electronic structures. The hopping integrals, which arise from the neighboring
and next-neighboring layers, cover the vertical and non-vertical interlayer atomic interactions. The band
structures exhibit three pairs of partially flat, sombrero-shaped and linear energy bands; the distinct
energy dispersions centered at the K point have been verified from the ARPES
experiments.\cite{NPGAsiaMat10;466,PRB88;155439,PRL98;206802} Specifically, the first pair just
overlaps/touches at $E_F$, leading to a sharp density of states (DOS). Such electronic wave functions are
ascribed to the special surface states due to the main contributions of carbon atoms in the outmost two
layers.\cite{RevModPhys81;109,PRB84;165404,PCCP17;26008} Although the ABC stacking is a semimetal with a
very low free carrier density, it is predicted to exhibit the low-frequency plasmom mode closely related
to the localized states.\cite{PRB98;041408} Furthermore, the novel momentum dependence of the pristine
plasmons is never observed in other stackings. On the other hand, the doping carriers in extrinsic
systems will strongly compete with the original surface states. It is expected to create dramatic changes
in the characteristics of e-h dampings and plasmon modes during the variation of $E_F$.

The AAB stacking is the direct combination of AA and AB stackings, as displayed in Fig. 1-1(e). Such
system has been successfully synthesized and experimentally observed by the distinct methods, e.g., the
mechanical exfoliation directly by a scalper or scotch tape,\cite{SurfSci601;498} the CVD growth on SiC
substrate\cite{PRB79;125411} and Ru(0001) surface,\cite{APL107;263101} the liquid-phase exfoliation of
natural graphite in N-methyl-2-pyrrolidone\cite{APL102;163111}, and the
STM.\cite{ACSNano7;1718,PRL108;205503,NatNanotech13;204,PRB48;17427,PRB86;085428,PRB50;1839,
JJAP52;035104}
In particular, the AAB stacking could be obtained by the rotating or horizontal shifting of the top
graphite layer along the armchair direction. That is to say, the stacking configuration is continuously
changed under the electrostatic modulation of STM.
Furthermore, the corresponding DOS is also measured, indicating a narrow energy gap in trilayer ABA
stacking.\cite{APL107;263101,SciRep6;33487} According to the first-principles method, the ground state
energies per unit cell of six carbon atoms in trilayer graphenes are evaluated for the our stacking
configurations. They are estimated as follows: $-$55.832866 eV, $-$55.857749 eV, $-$55.862386 eV and
$-$55.864039 eV  for AAA, AAB, ABA and ABC stackings, respectively.\cite{CRCPress;9781138556522} The
theoretical calculations predict that the AAB stacking is more stable than the AAA one, or the former
presents the more promising future in experimental syntheses. The lower-symmetry AAB stacking possess the
most complicated interlayer hopping integrals, in which this special property is clearly identified from
a consistent/detailed comparison between the tight-binding model and VASP calculations in the low-lying
energy bands.\cite{PCCP18;17597,Carbon94;619} For example, the AAB-stacked trilayer graphene exhibits
three pairs of energy bands with the oscillatory, sombrero-shaped and parabolic dispersions, in which the
first ones determine a very band gap of ${<10}$ meV. Of course, the semiconducting pristine system only
creates the inter-$\pi$-band e-h excitations, but not the low-frequency plamsons. The doping effects on
the Coulomb excitations are the first theoretical study, and the greatly diversified phenomena are
described in Chap. 7.

How to create the dramatic transitions of essential properties is one of the main-stream topics in
pristine graphenes. The continuous stacking configurations, which possess the high and low geometric
symmetries, could enrich and diversify the physical phenomena. The sliding bilayer graphene presents the
transformation between the highly symmetric stackings, being an ideal system for fully exploring the
electronic topological transitions. There are some experimental successful syntheses, such as, the
stacking boundaries including the relative shifts between neighboring graphene layers by the CVD
method,\cite{PNAS110;11256} the sliding of graphene flakes on graphene substrate/micrometer-size graphite
flakes initiated by the STM tip\cite{ACSNano7;1718,PRL108;205503}, and AFM tip\cite{NatNanotech13;204}
and the electrostatic-manipulation STM performed on a highly oriented pyrolytic graphite (HOPG)
surface.\cite{JJAP52;035104,PRB48;17427,PRB86;085428,PRB50;1839}
Specifically, the last method has generated a continuous and large-scaled movement of the top graphene
layer, so that the sliding bilayer graphene is expected to be achieved by this method.
The theoretical model calculations are focused on the
electronic,\cite{arXiv180510775v2,PRL100;036804,SciRep5;17490,JPCC119;10623,
PRB89;085426,PRB84;045404,PRB84;155410}
magnetic,\cite{SciRep4;7509}
optical,\cite{APE9;065103,RSCAdv4;63779}
transport\cite{APL103;173519,Carbon99;432,PRB88;115409} and
phonon \cite{ACSNano7;7151,PRL109;236604,PRB88;035428,PRL99;256802} properties.
For example, two low-lying isotropic Dirac cones are dramatically transformed into two pairs of parabolic
bands during the variation of AA$\to$AB.\cite{SciRep4;7509,JPCC119;10623} Furthermore, the sliding
bilayer graphene exhibits three kinds of Landau levels (LLs), the well-behaved, perturbed and undefined
LLs, and three magneto-optical selection rules of ${\Delta\,n = 0}$, 1 $\&$ 2 ($n$: the quantum number
for each LL).\cite{SciRep4;7509,IOPBook;978} During the continuous variation of stacking configuration,
energy bands present the serious distortions and free carrier density show the drastic
changes.\cite{SciRep4;7509,IOPBook;978} These are predicted to induce the novel Coulomb excitation
phenomena in pristine and extrinsic systems.

The external electric and magnetic fields are one of the critical factors in the creation of the diverse
electronic excitations. A uniform perpendicular electric field (${E_z\hat z}$) could be achieved by
applying a gate voltage on a layered graphene system. There are a lot of such experimental setups up to
now, being verified to have strong effects on transport properties and thus potential applications in
nano-electronic devices.\cite{MRE5;045804,AdvMat30;1705088,PCCP47;32602,PRB96;245410} $E_z$ creates the
distinct Coulomb potential energies on the layer-dependent sublattices. Apparently, energy dispersions
and band gap are drastically changed by $E_z$, in which the semi-metal-semiconductor transitions might
occur as $E_z$ varies\cite{NanoLett15;4429,Nature459;820}. The effects of $E_z$ on band structures are
also diversified by the stacking configurations; that is, the $E_z$-enriched energy bands are sensitive
to the AAA, ABA, ABC and AAB stackings (the detailed discussions on Chap. 9). From the current model
calculations of the modified RPA, a pristine $N$-layer AAA stacking has one acoustic plasmon mode and
${(N-1)}$ optical ones,\cite{PRB86;125434,AOP339;298} in which the former and the latter, respectively,
originate from the intraband and the interband electronic excitations. An electric field obviously
results in the charge transfer among the different graphene layers; therefore, it could modulate the
number, frequency and intensity of collective excitation modes, e.g., the emergence of new plasmon modes
and the decline of threshold plasmon frequency. As for the AB bilayer stacking, the low-frequency
plasmon, which corresponds to the intraband single-particle excitations, is generated by a sufficiently
high  $E_z$.\cite{SciRep3;1368} This greatly contrasts with a pristine system with any plasmon
mode.\cite{PRB74;085406} Such result directly illustrate that the $E_z$-induced oscillatory parabolic
bands, with the high DOSs, could create new plasmon modes, or the main features of band structures could
determine the single-particle and collective excitations. The similar effects are revealed in the
trilayer ABA and ABC stackings.\cite{arXiv:1803;10715,PRB98;041408} This work will cover the complete
results in the $E_z$-enriched excitation spectra of the ABA, ABC and AAB stackings.

A uniform magnetic field (${B_z\hat z}$) could flock together the neighboring electronic ${(k_x,k_y)}$
states and thus generate the highly-degenerate LLs and the special wave functions in the oscillatory
forms. Apparently, the magnetic quantization is directly reflected in magneto-electronic excitations,
belonging to a new topic of the Coulomb interactions,  The LL characteristics, which are thoroughly
explored by the generalized tight-binding
model,\cite{IOPBook;978,IOPChen,PCCP17;26008,CRCPress;9781138571068} cover the normal or abnormal spatial
probability distributions with the regular/irregular zero-point numbers,  the definition of quantum
number from the domination sublattice, the dependences of energy spectrum on $n$ and $B_z$, the
non-crossing/crossing/anti-crossing behaviors, and the specific magneto-optical selection
rules.\cite{IOPBook;978,CRCPress;9781138571068} They strongly depend on the number of layer and stacking
configuration, i.e., they are greatly diversified by the geometric factors. The charged particles under
${B_z\hat z}$ experience a transverse magnetic force. The cyclotron motions present rather strong
competitions with the longitudinal plasma oscillations due to the electron-electron interactions. This
critical mechanism is responsible for the unusual excitation phenomena.
Up to date, there are only few theoretical predictions on the magneto-plasmon modes of monolayer and
AA/AB bilayer graphenes.\cite{ACSNano5;1026,PRB89;165407,PRB94;205427} The direct combination of the
generalized tight-binding model and the modified RPA is a developed theoretical framework. In addition to
temperature, a magnetic field in monolayer graphene could drive a lot of discrete magneto-plasmons even
under the low-energy range, mainly owing to the inter-LL excitations. Their oscillation frequencies
exhibit the critical transferred momenta and the non-monotonous momentum dependence. Moreover, the
quantized magneto-plasmons and the 2D acoustic plasmon mode might coexist in the AA bilayer stacking, but
not the AB one. This difference lies in the stacking-dependent free carrier densities.

Graphite is one of the most extensively investigated materials experimentally and theoretically, being
much more than 100 years. This layered system is very suitable/ideal for studying the diverse 3D and 2D
physical phenomena; furthermore, it possesses a lot of up-to-date applications. Graphite crystals consist
of a series of stacked graphene plane. Generally, there exist three kinds of stacking configurations in
the layered graphites and compounds, namely AA, AB and ABC stackings. Simple hexagonal, Bernal and
rhombohedral graphites exhibit the unusual essential properties as a result of the honeycomb lattice and
stacking sequence. Among them, the AB-stacked graphite is predicted to to be the most stable system,
according to the first-principles calculations on the ground state energy.\cite{Carbon32;289}  Natural
graphite is composed of the dominating AB stacking and the partial ABC
one.\cite{PRSLSA106;749,PRSLSA181;101} The AA-stacked graphite, which possesses the highest-symmetry
crystal structure, does not survive in nature. The periodical AA stacking is frequently observed in the
Li-intercalation graphite compounds.\cite{AdvPhys30;139} Simple hexagonal graphite has been successfully
synthesized by using the dc plasma in hydrogen-methane mixtures.\cite{JCP129;234709}
For the AA-. AB-  and ABC-stacked graphites, the first/third system owns the highest/lowest 3D free
carrier density.\cite{CRCPress;9781138571068} Their low-energy band structures, respectively, present the
${(k_x,k_y)}$-plane Dirac cones, the monolayer/bilayer characteristics at the H/K point, and the spiral
Dirac-cone structure.\cite{CRCPress;9781138571068} The ARPES measurements on Bernal graphite have
verified the linear and parabolic energy dispersions near the H and K
valleys.\cite{PRL100;037601,NatPhys2;595,ASS354;229,PRB79;125438} The theoretical studies of Coulomb
excitations are focused on the free-carrier-induced, $\pi$, and ${\pi\,+\sigma}$ plasmons, in which they
are, respectively, revealed in the distinct frequency ranges of ${\omega_p\,\sim}$0.1 eV, 5 eV; 20 eV.
The current investigation is the first theoretical study on whether the low-frequency plasmons exist in
rhombohedral graphite.\cite{JPSJ69;3781,JPSJ70;897,JPSJ81;104703} The low-frequency plasmons belong to
the optical mode because of the 3D bare Coulomb potential. Apparently, dimensionality and stacking
configuration/interlayer hopping integrals play critical roles in determining its characteristics, the
existence and momentum $\&$ temperature dependence. On the experimental side, the reflection energy loss
spectroscopy (REELS; details in Chap. 2), with a very high energy resolution, has been utilized to
thoroughly examine the $T$-dependent low-frequency optical plasmon in Bernal graphite. Furthermore, both
REELS and TEELS (transmission energy loss spectroscopy) could accurately identify the middle-frequency
$\pi$ plasmon modes.

Carbon nanotubes, with the very strong $\sigma$ bondings, are first discovered by Iijima using an
arc-discharge evaporation method in 1991.\cite{Nature354;56,Nature363;603}
Each nanotube is a hollow cylinder, so that it could be regarded as a rolled-up graphitic sheet in the
cylindrical form. Its geometric structure is characterized by a primitive lattice vector (${\bf R}$) of
monolayer graphene (details in Chap. 12). Both radius ($r$) and chiral angle ($\theta$), which correspond
to ${\bf R}$, would dominate the essential physical properties.
From the theoretical calculations of the VASP and tight-binding
model,\cite{PRB67;045405,APL60;2204,PRL72;1878} there are three kinds of carbon nanotubes, namely,
metallic, narrow-gap and middle-gap ones, being determined by ${(r,\theta\,)}$. Only armchair nanotubes
belong to 1D metals, since DOSs due to the linear bands are finite at $E_F$. The
${(r,\theta\,)}$-dependent energy gaps are thoroughly verified from the STS measurements on the
low-energy DOSs.\cite{Nature391;59} Any cylindrical nanotubes present the semiconductor-metal transitions
during the variation of a uniform axial magnetic field The double degeneracy of electronic states are
destroyed under this field. Furthermore, the periodical oscillations of essential properties under a flux
quantum ${\phi_0\,=hc/e}$) is the so-called Aharonov-Bohm effect, as clearly observed in
optical\cite{Science304;1129} and transport properties.\cite{Nature397;673} As to a lot of 1D energy
subbands of a cylindrical carbon nanotube, they are defined by the angular momenta ($J$$^{,}$s) and axial
wave vectors ($k_y$$^{,}$s). The e-e Coulomb interactions would satisfy the conservation of the
transferred angular momentum and axial momentum
($L,q$).\cite{JPSJ66;757,PRB47;6617,PRB56;4996,PRB76;115422} Specifically, both single-particle and
collective excitations possess the $L$-decoupled modes, i.e., there exist many intra-$\pi$-band and
inter-$\pi$-band transitions. The previous theoretical studies show that the free carriers in a metallic
armchair nanotube create a 1D $L=0$ acoustic plasmon mode with a specific momentum dependence. All the
carbon nanotubes exhibit several inter-$\pi$-band plasmons of ${\omega_p\sim\,1-4}$ eV and one $\pi$
plasmon mode of ${\omega_p\,>5}$ eV, being consistent with the experimental
measurements.\cite{PRL100;196803}

The monoelement IV-group condensed-matter systems have attracted a lot of
experimental\cite{ProgSurfSci90;1,NanoLett15;2510,NatMater14;1020} and theoretical
studies,\cite{PRL109;055502,PRB84;195430,NJP16;125002,PRB94;045410,PRB94;205427} especially for those
combined with 2D $\&$ 3D structures. The emergent 2D materials are excellent candidates in exploring the
unique physical phenomena, such as, the Dirac-cone band structure or the multi-constant-energy
loops,\cite{PRB85;075423,PRL111;136804,JPCC119;11896} the magnetically quantized
LLs,\cite{PRL110;197402,PRB91;035423} the ultrahigh carrier
mobility,\cite{Naturetech10;227,SciRep5;14815} the novel quantum Hall
effects,\cite{PRB90;235423,PRL109;055502} and the optical selection
rules.\cite{PRL110;197402,PRB88;085434} These systems are expected to present high potentials in the
near-future technological applications.\cite{NatMater16;163,SurSciRep67;1} Since the successful
exfoliation of few-layer graphenes in 2004,\cite{Science306;666} silicene germanene and tinene, are
respectively, grown on the distinct substrate surfaces, e.g., Si on Ag(111), Ir(111) $\&$ ZrB$_{2}$
surfaces,\cite{PRL108;155501,NanoLett13;685,PRL108;245501} Ge on Pt(111), Au(111) $\&$ Al(111)
surfaces,\cite{AdvMater26;4820,NJP16;095002,2DMater4;031005,NanoLett15;2510} and Sn on Bi$_2$Te$_3$
surface. The latter three systems possess the buckled honeycomb structures, in which the strong
competition between the sp$^2$ and sp$^3$ bondings accounts for the optimal geometries. However, graphene
is a hexagonal plane. Their spin-orbit couplings (SOCs) are significant and much stronger than those of
pure carbon systems. These two characteristics will dominate the essential physical properties. From the
VASP\cite{JPCC119;11896} and tight-binding model calculations,\cite{PRB84;195430} the low-lying
electronic structures of monolayer silicene and germanene appear at the K/K$^\prime$ valley; furthermore,
they are mainly determined by the outermost 3$p_z$/4$p_z$ orbitals even in the mixing of two kinds of
chemical bondings. The non-negligible SOCs create the separated Dirac-cone structures, with narrow energy
gaps, e.g., ${E_g\sim\,7.9}$ meV for silicene and ${E_g\sim\,93}$ meV for germanene for the model
predictions.\cite{JPCC119;11896,PRB84;195430} Compared with monolayer graphene, it is relatively easy to
reveal the anisotropic energy dispersions as a result of the smaller intralayer hopping integrals. The
application of a uniform perpendicular electric field further induces the splitting of spin-related
energy bands.\cite{NJP16;095002,2DMater4;031005,NJP14;033003,RSCAdv5;51912} Also, this $E_z$-field causes
energy gap to change from the finite to zero values. On the other hand, monolayer tine exhibit the very
strong multi-orbital hybridizations and SOCs,\cite{PRB94;045410} so that the outside four orbitals
${(5s,5p_x,5p_y,5p_z)}$ need to be considered in the low-energy band structures. The calculated
electronic energy spectra could be verified from the ARPES
measurements,\cite{PRL98;206802,PRB88;075406,PRB88;155439,NPGAsiaMat10;466,
NatMater23;887,NanoLett17;1564,
PRL100;037601,ASS354;229,NatPhys2;595,PRB79;125438}
as done for few-layer graphenes and graphites.

For monolayer silicene and germmanene, there are some theoretical studies on Coulomb excitations/daeay
rates.\cite{PRB89;195410,SciRep7;40600,NJP16;125002,RSCAdv5;51912,PRB94;205427} These two systems are
different from monolayer graphene in certain many-particle properties, mainly owing to the existence of
SOCs and buckled structures. Energy gap, electric field, magnetic field and doping would induce the
diversities of  excitation phenomena, in which the diverse momentum-frequency phase diagrams cover the
various single-particle excitation boundaries and  plasmon modes, e.g., four kinds of $E_F$-, SOC- and
$E_z$-dependent plasmon modes in germanene.\cite{PRB94;205427} Such excitations might become the
effective deexcitation channels, when the occupied electrons/holes are excited into the unoccupied states
under the perturbation of an incident electron beam or an electromagnetic field. That is to say, the
excited electrons or holes in conduction/valence bands could further decay by the inelastic Coulomb
scatterings. The decay rates of the excited states have been explored by the screening exchange exchange
using the Matsubara Green$^{,}$s functions. The dynamic Coulomb responses from the valence and conduction
electrons are taken into account, simultaneously.\cite{PRB89;195410,
SciRep7;40600,NJP16;125002,RSCAdv5;51912,PRB94;205427} The decay processes and their dependence on the
wave vector, valence/conduction states, and Fermi energy/doping density will be investigated in detail. A
comparison with monolayer graphene is also made. The current work indicates that the intraband $\&$
interband single-particle excitations, and the distinct plasmon modes are responsible for the
deexcitation behaviors. The rich and unique Coulomb decay rates appear as a consequence of the
oscillatory energy dependence, the strong anisotropy on wave vectors, the non-equivalent valence and
conduction states/Dirac points, and the similarity with 2D electron gas for the low-energy conduction
electrons and holes. The predicted Coulomb decay rates could be directly examined from the
high-resolution ARPES measurements by the energy widths of quasiparticle states at low
temperature.\cite{NatMater23;887,NanoLett17;1564,NatPhys2;595,ASS354;229,PRB79;125438}

This book is focused on the recent progresses of graphene-related systems in Coulomb
excitations/deexcitations under the electron-electron interactions. The whole content is organized as
follows. Chapter 2 covers the developed theoretical framework and the detailed experimental techniques.
The bare and screened response functions are derived in the analytic forms, especially for those in
few-layer graphenes using the modified RPA. The geometric symmetry, layer number, stacking configuration,
dimension; electric and magnetic fields are taken into account simultaneously. How to accurately measure
the energy loss spectra is thoroughly discussed for the high-resolution REELS and TEELS. The previous
experimental measurements on carbon-sp$^2$ condensed-matter systems provide very useful information on
the geometry-enriched electronic excitation behaviors. As for the Coulomb decay rates in monolayer
graphene, silicene and germanene, they are calculated from the self-energy method of Matsubara$^{,}$s
Green functions. The inelastic Coulomb scatterings from the various valence and conduction states play
the critical roles.

The focuses in Chap. 3 are the effects of the critical factors, temperature and doping, on creating the
intraband single-particle excitations and low-frequency 2D acoustic plasmons. The low-frequency analytic
formula of the polarization function could be obtained in the vicinity of the K/K$^\prime$. Also, the
pure Coulomb coupling effects are explored for a double-layer system with a sufficiently large interlayer
distance. The composite effects, which are due to the interlayer hopping integrals and the interlayer
Coulomb interactions in few-layer graphene systems, will create the diverse excitation phenomena in the
spectra of (frequency,momentum)-phase diagrams. Their systematic studies of the AAA-, ABA-, ABC- and AAB-
graphenes are, respectively, conducted on Chaps. 4, 5, 6 and 7. In addition, Chap. 8, as fully
investigated for the sliding bilayer graphene, clearly illustrates the continuous transformation of the
geometric symmetry and thus the dramatic changes in electronic properties and Coulomb excitations. The
layer-dependent Coulomb potentials, being induced by a uniform perpendicular electric field, could
greatly diversify band structures and excitation behaviors, as clearly indicated in Chap. 9. The magnetic
quantization is elaborated in chap. 10; we define a vetoer-potential-dependent Peierls phase/period in
the hopping integrals/the real crystal.\cite{CRCPress;9781138571068} There exist the new excitation
channels, the inter-LL single-particle and collective excitations, which display unusual momentum
dependences. The generalized tight-binding model and the modified RPA are combined to fully comprehend
the AA and AB bilayer stackings. Also, the strong competitions among the longitudinal Coulomb
interactions, the transverse magnetic forces, and the stacking configurations are discussed in detail.


The dimension- and geometry-enriched electronic excitations are, respectively, explored in Chaps. 11 and
12 for 3D graphites and 1D carbon nanotubes. The simple hexagonal, Bernal and rhombohedral graphites are
in sharp contrast with one another for the low-lying band structures, covering energy dispersions,
isotropic/anisotropic behaviors, and free electron/hole densities. How to directly reflect in the
low-frequency electronic excitations is worthy of a systematic investigation. As for each carbon
nanotube, the cylindrical symmetry creates a lot of the angular-momentum-decoupled excitation modes,
being absent in other condensed-matter systems. A close relationship between 1D band structures and
excitation spectra is proposed to fully understand the unusual Coulomb excitations. The important
differences among 1D, 2D and 3D graphene-related systems require a detailed discussion. Chapters 13 and
14, respectively, correspond to electronic excitations and Coulomb decay rates in the emergent monolayer
silicene/germanene, in which the significant SOCs and buckled structures are the critical factors in
determining the diversified Coulomb excitations and decay rates, or  distinguishing from monolayer
graphene. Finally, chap. 15 includes concluding remarks and future perspectives.

\section{Theories for electronic excitations in layered graphenes, 3D graphites and 1D carbon nanotubes;
experimental equipments }

We will develop the theoretical modes for the dielectric responses of graphene-related systems with
various dimensionalities. The electron-electron Coulomb interactions due to the $\pi$ electrons induce
the dynamic and static charge screenings within the middle frequency (${\omega\le\,10}$ eV). Within the
linear response, the random-phase approximation under the external electric and magnetic fields is
modified to satisfy the layered structures/the geometric symmetries. The energy loss function, which is
directly related to the experimental measurements, is very useful in exploring the unusual electronic
excitations in each system. It can be further defined by an analytical formula. In general, there are two
kinds of experimental equipments in examining the theoretical predictions. The significant
characteristics of the distinct dimensional systems  in the measured energy loss spectra are discussed
thoroughly. Moreover, the inelastic Coulomb decay rates are also investigated in detail.

\subsection{Dielectric Functions of layered graphenes}

When monolayer graphene is present in an external Coulomb potential, the $\pi$ electrons due to the
${2p_z}$ orbitals will effectively screen this perturbation. The charge redistribution directly reflects
the dynamic/static carrier screening and thus creates to the induced potential. Within the linear
response, the dimensionless dielectric function is defined as the ratio between the bare potential and
the effective potential
\begin{equation}
\epsilon(\mathbf{q},\omega)=\lim_{V^{ex}\rightarrow
0}\frac{V^{ex}(\mathbf{q},\omega)}{V^{eff}(\mathbf{q},\omega)}.
\end{equation}
It can also be characterized by the charge densities and the longitudinal electric fields, namely,
$\rho^{ex}/\rho^{tot}$ and ${D_l/E_l}$. By the momentum-dependent Poisson equations and the
self-consistent-field approach, the induced Coulomb potential is the product of the bare Coulomb
potential and the induced charged density, in which the latter is proportional to the effective Coulomb
potential, and the coefficient is bare response function ($P$) under the linear response. As a result,
the dielectric function is given by
\begin{equation}
{\epsilon\,(q,\phi\,,\omega\,)=\epsilon_0\,-V_qP(q,\phi\,,\omega\,)}\text{,}
\end{equation}

where
\begin{equation}
\begin{split}
P(q,\phi\,,\omega\,)&=\sum_{h,h^{\prime}=c,v} \langle \mathbf{k};h|e^{-i\mathbf{q}\cdot
\mathbf{r}}|\mathbf{k+q};h^{\prime}\rangle
\times \frac{f(E^{h^{\prime}}(\mathbf{k+q}))-f(E^{h}(\mathbf{k}))}
{E^{h^{\prime}}(\mathbf{k+q})-E^{h}(\mathbf{k})-(\omega+i\Gamma)}\text{.}
\end{split}
\end{equation}


In general, a 2D system has an electronic state expressed by (${k_x, k_y; h}$). ${h=v/c}$ corresponds to
valence/conduction state. $\epsilon_{0}$(=2.4) is the background dielectric constant due to the
high-energy $\sigma$-electron excitations. ${V_q=2\pi\,e^2/q}$ is the 2D bare Coulomb potential of the 2D
electron gas. The band-structure effect on the bare Coulomb interactions are included in the bare
response function by the square of the inner product between the initial and final states in the momentum
transfer (the first term inside the integration on the first Brillouin zone of Eq. (3)). $f$ is the
Fermi-Dirac distribution function, and Eq. (2) is suitable under any temperatures in intrinsic and
extrinsic systems. $\Gamma$ is the broadening phenomenological parameter arising from the various
deexcitation channels, depending on the frequency range of electronic excitations. The transferred
momentum and frequency are conserved during the electron-electron Coulomb interactions; that is, they are
necessary in describing the electronic excitations, the single- and many-particle excitations. In
inelastic experimental measurements, the energy loss spectra are associated with ${\bf q}$ and $\omega$.
The direction of the former is $\phi$ between ${\bf q}$ and KM, and ${0^\circ\le\phi\le\,30^\circ}$ is
sufficient because of the hexagonal symmetry. The dielectric function in Eqs. (2) and (3) is similar for
monolayer silicene and germanene with buckled honeycomb lattices, while electronic states are modified by
the significant spin-orbital interactions (details in Chapter 13).

The dielectric function could be used to explore the effective Coulomb interactions between two charges
and thus understand the screening length. First, we calculate the static dielectric functions which
depend on the Fermi level of monolayer graphene. For an intrinsic (extrinsic) system, the Fermi level is
located at  the Dirac point (the conduction/valence cone), so monolayer graphene is a zero-gap
semiconductor (a metal with the free electron/hole density roughly proportional to the square of the
Fermi momentum). As a result, the dielectric function is finite/divergent under the long wavelength limit
(${q\to\,0}$) for an intrinsic/extrinsic graphene. Second, the momentum-dependent effective Coulomb
potential is transformed into the real-space electron-electron interaction by the standard 2D Fourier
transform. The long-range behavior, the effective interaction inversely proportional to the distance, is
deduced to remain  in an intrinsic graphene. However, for an extrinsic graphene, the electron-electron
interaction close to the charged impurity will  decline quickly and exhibit an effective screening length
closely related to the free carrier density. Third, the similar Fourier transform is done for the
real-space induced charge density with the well-known Friedel oscillations associated with the
dimensionality and Fermi momentum.

The dielectric responses of $N$-layer graphens become more complicated, compared with monolayer system.
There exist the perturbed Coulomb potentials and the induced charges arising from all the layers; that
is, the external and induced Coulomb potentials due to each layer need to be taken into consideration
simultaneously. The intralayer $\&$ interlayer hopping integrals and the intralayer $\&$ interlayer
Coulomb interactions are covered in the modified RPA. The full band structure can provide the exact and
reliable electronic excitations, in which the rich and unique ($\mathbf{q},\omega$)-phase diagrams are
very sensitive to the stacking configuration and the number of layers.

The incident electron beam is assumed to be uniform on each layer, so the $\pi$ electrons on the distinct
layers experience the similar bare Coulomb potentials. Such carriers exhibit the dielectric screening
closely related to two different layers/the same layer. Specifically, the excited electron and hole in
each excitation pair, which is due to the Coulomb perturbation, frequently occur on distinct layers. By
the Dyson equation, the effective Coulomb potential for two electrons on the $l$-th and $l^\prime$-th
layers is expressed as

\begin{figure}[htbp]
\center
\rotatebox{0} {\includegraphics[width=17cm]{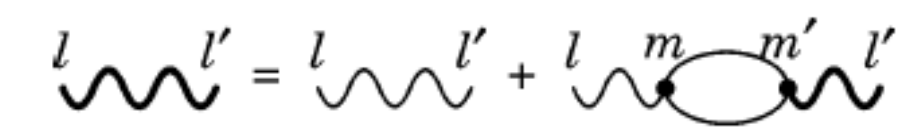}}
\caption{The Feynman diagram of the Coulomb excitations between two layers under the modified RPA.}
\label{Figure}
\end{figure}

\begin{equation}
\epsilon_{0}V^{eff}_{ll^{\prime}}(\mathbf{q},\omega)=V_{ll^{\prime}}(\mathbf{q})+
\sum\limits_{mm^{\prime}}V_{lm}(\mathbf{q})P^{(1)}_{mm^{\prime}}(\mathbf{q},\omega)
V^{eff}_{m^{\prime}l^{\prime}}(\mathbf{q},\omega)\text{.}
\end{equation}%

Equation (4) clearly reveals the bilayer-dependent effective Coulomb potential as the ${N\times\,N}$
matrix form, in which the Coulomb potential, induced charge density and response function are described
by any two layers. The first term is also useful in understanding the Coulomb decay rates in layered
systems. Specifically, the bilayer-created response function is

\begin{equation}
\begin{array}{l}
P_{mm^{\prime}}(\mathbf{q},\omega)=2\sum\limits_{k}\sum\limits_{nn^{\prime}}\sum\limits_{h,h^{\prime}=c,v}
\biggl(\sum\limits_{i}u^{h}_{nmi}(\mathbf{k})u^{h^{\prime}*}_{n^{\prime}m^{\prime}i}(\mathbf{k+q})\biggr)\\
\times\biggl(\sum\limits_{i^{\prime}}u^{h*}_{nmi^{\prime}}(\mathbf{k})
u^{h^{\prime}}_{n^{\prime}m^{\prime}i^{\prime}}
(\mathbf{k+q})\biggr)\times\frac{f(E^{h}_{n}(\mathbf{k}))-f(E^{h^{\prime}}_{n^{\prime}}(\mathbf{k+q}))}
{E^{h}_{n}(\mathbf{k})-E^{h^{\prime}}_{n^{\prime}}(\mathbf{k+q})
+\hbar\omega+i\Gamma}
\text{.}
\end{array}
\end{equation}%

Any electronic states, which agree with the conservations of the transferred momentum and frequency, can
be decomposed into the layer-dependent contributions by analyzing their wave functions, and so does the
response function. We have included all the significant intralayer and interlayer hopping integrals in
Eq. (5). This decomposition concept is critically important to match with the layer-dependent Coulomb
potential in Eq. (4), so that the modified random-phase approximation could be generalized to the layered
systems even in the presence of the external electric and magnetic fields. Moreover, the layer-dependent
dielectric function becomes a tensor form:

\begin{equation}
\epsilon_{ll^{\prime}}(\mathbf{q},\omega)=\epsilon_{0}\delta_{ll^{\prime}}-
\sum\limits_{m}V_{lm}(\mathbf{q})P_{m,l^{\prime}}(\mathbf{q},\omega)\text{.}
\end{equation}%

The zero points of the dielectric function tensor are available in understanding the plasmon modes, while
the spectral intensities of the collective and single-particle excitations are absent. The definition of
the energy loss function is necessary in the further calculated formulas. The effective Coulomb potential
directly links with the bare one through the following relationship

\begin{equation}
\sum_{l^{\prime\prime}}\epsilon_{ll^{\prime\prime}}(\mathbf{q},\omega)V^{eff}_{l^{\prime\prime}l^{\prime}}
(\mathbf{q},\omega)=V_{ll^{\prime}}(\mathbf{q})\text{.}
\end{equation}%

The inelastic scattering rate, which the probing electrons transfer the specific momentum and frequency
${(\bf q\,,\omega\,)}$ to the $N$-layer 2D materials, is delicately  evaluated from the Born
approximation.$^{Rs}$ It is used to defined the dimensionless energy loss function:

\begin{equation}
\begin{split}
\mathbf{Im}[-1/\epsilon]&\equiv\sum\limits_{l}\mathbf{Im}\biggl[-V^{eff}_{ll}(\mathbf{q},\omega) \biggr]
/\biggl(\sum\limits_{lm}V_{lm}(q)/N\biggl)\text{.}
\end{split}
\end{equation}%

The denominator is the average of all the external Coulomb potentials on the different layers. Equation
(8) can be applied to any emergent 2D systems, such as, the layered graphene, silicene, germanene,
tinene, phosphorene, antimonene and  bismuthene (the group-IV and group-V 2D materials). It is the
screened response function responsible for the experimental inelastic energy loss spectra. The
dimensionless loss function is useful in exploring the various plasmon modes in the specific systems, and
the imaginary of the bare response function describes the single-particle electron-hole excitations.
All the equations developed in this section is suitable for a layered condensed-matter system under a
uniform perpendicular electric field. It only needs to  modify the band-structure changes due to the
layer-dependent Coulomb site energies.

\subsection{AA-, AB- and ABC-stacked graphites}

The bulk graphites possess the infinite graphene layers, so that their energy bands have an extra wave
vector along the $k_z$-direction. Electronic states are described by ${(k_x,k_y,k_z)}$ within the first
Brillouin zones. Energy dispersions are dominated by the honeycomb lattice on the ${(x,y)}$ plane,
stacking configuration; intralayer and interlayer hopping integrals. All the graphites are semi-metals
because of the interlayer van der Waals interactions. However, the AA-stacked graphite (ABC-stacked one)
exhibits the largest (smallest) overlap between the valence and conduction bands and thus the highest
(lowest) free electron and hole densities, directly reflecting the geometric symmetry. Band structures
and free carrier densities are quite different for three kinds of graphites, and electronic excitations
are expected to behave so (details in Chap. 11).\cite{JPSJ69;3781,JPSJ70;897,JPSJ81;104703} For example,
the low-frequency plasmon due to the interlayer atomic interactions are very sensitive to AA, AB or ABC
stacking.

The 3D transferred momentum (${q_x,q_y,q_z}$) in graphites is conserved during the electron-electron
Coulomb intercations, as observed in 3D electron gas. The analytic form of the dielectric function, which
is similar for any graphites, is directly evaluated from the RPA

\begin{equation}
\begin{split}
\epsilon(q_{x},q_{y},q_{z},\omega\,)&=\epsilon_{0}-\sum_{h,h^{\prime}=c,v} \int_{1st
BZ}\frac{e^{2}d^{2}\mathbf{k}_{\parallel}dk_{z}}{q^{2}\pi^{2}}|\langle
\mathbf{k}_{\parallel}+\mathbf{q}_{\parallel},k_{z}+q_{z};h^{\prime}|e^{i\mathbf{q\cdot
r}}|\mathbf{k}_{\parallel},k_{z};h\rangle|^{2}\\
&\times \frac{f(E^{h^{\prime}}(\mathbf{k}_{\parallel}+\mathbf{q}_{\parallel},k_{z}+q_{z}))-
f(E^{h}(\mathbf{k}_{\parallel},k_{z}))}
{E^{h^{\prime}}(\mathbf{k}_{\parallel}+\mathbf{q}_{\parallel},k_{z}+q_{z})-
E^{h}(\mathbf{k}_{\parallel},k_{z})-(\omega+i\Gamma)}\text{.}
\end{split}
\end{equation}

The $k_z$-integration of the first Brillouin zone is distinct in three kinds of stackings configurations.
There exist the low-frequency and $\pi$ plasmons. The former could survive at small transfer momenta,
while it might be difficult to observe it at large ones. Under this case, the anisotropic dependence on
the ${(q_x,q_y)}-$ plane is negligible; that is, ${\bf q\,=[\bf q_{\parallel}\,,q_z]}$. Since the
low-lying energy bands are almost isotropic near the K/K$^\prime$ point, the 3D integration could be
reduced to the 2D integration, i.e., ${\int_{1stBZ}\,d^2{\bf k}dk_z}$
$\to$${\int_{1stBZ}\,2\pi\,k_{\parallel}dk_{\parallel}dk_z}$.

Both 3D graphite and 2D monolayer graphene have the similar dielectric functions in Eqs. (8) and (2),
while their electronic excitations quite differ from each other. The dimensionality and band structure
are responsible for the significant differences. Under the long wavelength limit, the bare Coulomb
potentials, respectively, approach to ${1/q^2}$ and ${1/q}$ for the former and the latter. The stronger
Coulomb potential in graphite clearly indicates that it is relatively easily to observe the low-frequency
plasmon modes due to free carriers and the screening charge distributions. Moreover, graphites possess
the extra $k_z$-dependent energy dispersions, compared with that of graphene. The larger overlap of
valence and conduction bands arising from the interlayer hopping integrals results in free electrons and
holes simultaneously. These are useful in understanding why three kinds of graphites exhibit the rich and
unique ${({\bf q},\omega\,)}$-phase diagrams (Chap. 11)

\subsection{Carbon Nanotubes}

Electronic states in a cylindrical carbon nanotube, with the nanoscaled radius ($r$), are characterized
by the longitudinal wave vector ($k_y$) and the azimuthal angular momentum (${k_x=J/r}$, $J=1, 2,...,
N_{u/2}$, $N_{u}$) the atom number in a primitive cell; discussed in Chap. 12). As a result of the
cylindrical symmetry, the transferred longitudinal momentum and transverse angular momentum are conserved
during the electron-electron Coulomb interactions. Electronic excitations are well defined by ${(q,L)}$,
and so does the dielectric response. To within RPA, the dielectric function of a single-walled carbon
nanotube, which includes all the intra- and inter-$\pi$-band excitations at any temperatures, is

\begin{equation}
\begin{split}
\epsilon^{\prime}(q,L,\omega+i\Gamma)=\epsilon_{0}+2\sum_{J} \int_{1st BZ}\frac{dk_{y}}{(2\pi)^{2}}
\frac{2\omega_{vc}(J,k_{y};q,L)}{(\omega_{vc}(J,k_{y};q,L))^{2}-(\omega+i\Gamma)^{2}}\\
\times V(q,L)|\langle\ J+L,k_{y}+q;h^{\prime}|e^{iqy}e^{iL\phi^{\prime}}|J,k_{y};h\rangle|^{2}\text{,}
\end{split}
\end{equation}
where
\begin{equation}
\begin{split}
|\langle\ J+L,k_{y}+q;h^{\prime}|e^{iqy}e^{iL\phi^{\prime}}|J,k_{y};h\rangle|\\
=\frac{1}{4}\{1+[q^{2}+(L/r)^{2}]/36\}^{-6}\\
\times \biggl|
1-\frac{H_{12}(J+L,k_{y}+q)H^{*}_{12}(J,k_{y})}{|H_{12}(J+L,k_{y}+q)H^{*}_{12}(J,k_{y})|}\biggr|^{2}\text{.}
\end{split}
\end{equation}

$\omega_{vc}(J,k_{y};q,L)={E^{h^\prime}(J+L,k_y+q)}$-${E^{h}(J,k_y)}$ is the excitation energy between
the final and initial states. ${\epsilon_0\,-2.4}$ is identical to that in monolayer graphene.
${V(q,L)=4\pi\,e^2I_L(qr)K_L(qr)}$ is the bare Coulomb potential of an electron gas in a cylindrical
tubule.\cite{PRB47;6617,PRB50;17744} ${I_L}$ ($K_L$) is the first (second) kind of modified Bessel
function of the order $L$. The Coulomb interaction in a carbon nanotube is characterized by the second
term in the integrand of Eq. (2a), in which Eq. (2b) corresponds to the Coulomb matrix element. The
wavefunctions of energy bands greatly modify the Coulomb interactions and produce noticeable effects on
the physical properties of carbon nanotubes, e.g., EELS and impurity screenings.\cite{PRB56;4996}
Electronic excitations are expected to strong depend on the metallic and semiconducting behavior, being
sensitive to the radius and chiral angle of a cylindrical carbon nanotube.

A single-walled carbon nanotube is similar to monolayer graphene in geometric structures, but the former
needs to satisfy the periodical boundary condition. As a result, the important differences are revealed
in in Coulomb excitations, as clearly indicated the dielectric functions in Eq. (10) and (2). The 1D and
2D bare Coulomb potentials are divergent in the logarithmic and linear forms. In general, the 1D and 2D
parabolic (linear) bands DOSs exhibit the square-root divergent form and the shoulder structure (the
shoulder structure and the V-shape form). Moreover, carbon nanotubes possess the L-decoupled
single-particle excitations and plasmon modes, mainly owing to the cylindrical symmetry. It is thus
expected to have the $L$-dependent diverse phase diagrams. e.g., the $L$-decoupled inter-$\pi$-band
plasmons (Chap.12).

For a multi-walled carbon nanotube, there are complicated dynamic/static charge screenings from the
different layers. The layer-dependent Dyson equation and polarization function, as done for layered
graphenes, are available in exploring excitation spectra due to the multi-walled
systems.\cite{PRB76;115422} The Coulomb excitations are greatly diversified by the relative stacking,
layer number and chiral angle of coaxial carbon nanotubes. For example, the (5,5)-(10,10) bilayer
nanotube has three kinds of rotational symmetries, namely, C$_5$, D$_{5h}$ and S$_5$, in which they
create the diverse ${(q,L)}$-phase diagram with the distinct plasmon modes and electron-hole excitations
(Landau dampings).

\subsection{Electron excitations under a uniform perpendicular magnetic field}

Monolayer graphene consists of two sublattices of A and B atoms with a C-C bond length of b=1.42 ${\AA}$.
Only the hopping integral between two nearest-neighbor atoms, $\gamma_{0}=2.598$eV, is used in
the tight-binding model. Based on the two tight-binding functions of the periodic ${2p_z}$ orbitals, the
zero-field Hamiltonian is a ${2\times\,2}$ Hermitian matrix. Monolayer system exits in a perpendicular
uniform magnetic field $\mathbf{B}=B_{z}\hat{z}$. The magnetic flux through a hexagon is
$\Phi=(3\sqrt{3}b^{2}B_{z}/2)/\phi_{0}$ where the flux quantum $\phi_{0}=hc/e$ ${4,14\times\,10^{-15}}$
T$\cdot$m$^2$. The vector potential, being chosen as $\mathbf{A}=B_{z}x\widehat{y}$, creates a new
periodicity along the armchair direction, as clearly indicated in Fig. 2-1. The magnetic Hamiltonian
matrix element could be obtained by multiplication of the zero-field Hamiltonian matrix element by a
Peierls.\cite{CRCPress;9781138571068} Such phase is assumed to have an integer period $R_{B}=1/\Phi$, so
that the enlarged rectangular unit cell contains $4R_{B}$ carbon atoms and the magnetic Hamiltonian
becomes a $4R_{B}\times 4R_{B}$ Hermitian matrix. The first Brillouin zone is greatly reduced along
${\hat k_x}$. The nearest-neighbor hopping integral related to the extra position-dependent Peierls phase
is changed into

\begin{equation}
\begin{split}
|\langle\ B_{k}|H|A_{j}\rangle|&=\gamma_{0}\exp[i\mathbf{k}\cdot(R_{A_{j}}-R_{B_{k}})]\\
&\times \exp\biggl( i2\pi\phi_{0}\int_{R_{A_{j}}}^{R_{B_{k}}}\mathbf{A}d\mathbf{r}\biggl)\\
&=\gamma_{0}t_{j}\delta_{j,k}+\gamma_{0}s\delta_{j,k+1}
\text{,}
\end{split}
\end{equation}

where $t_{j}=\exp\{i[-k_{x}b/2-k_{y}(\sqrt{3}b)/2+\pi\Phi(j-1+1/6)]\}
+\exp\{i[-k_{x}b/2+k_{y}(\sqrt{3}b)/2-\pi\Phi(j-1+1/6)]\}$ and $s=\exp[i(-k_{x}b)]$.
The Hamiltonian dimension is 32000 under the magnetic field strength of ${B_z=10}$ T. The huge magnetic
Hamiltonian could be efficiently solved using the band-like matrix by the rearrangement of the $R_B$
tight-binding functions (details in Refs.\cite{IOPBook;978,IOPChen}). Monolayer graphene exhibits the
highly degenerate Landau levels and the well-behaved wave functions with spatial symmetries, in which the
significant quantum number, $n$/$m$, is determined by the number of zero points in the latter.

Magneto-electronic excitations are characterized by the transferred momentum q and the excitation
frequency $\omega$, which determine the longitudinal dielectric function. They are independent of the
direction of the momentum transfer, since the LLs possess the isotropic
characteristics.\cite{ACSNano5;1026,PRB89;165407} The magnetic dielectric function and bare response
function have the same forms identical to Eqs. (2)  and (3), respectively. In the presence of the
magnetic field, electronic states become fully quantized. The summation in Eq. (3) corresponds to all
possible single-particle transitions between Landau states $|m\rangle$ and $|n\rangle$. The response
function is now expressed as

\begin{equation}
\begin{split}
P(\mathbf{q},\omega)=\frac{1}{3bR_{B}\pi}\sum_{n,m;k}|\langle n;\mathbf{k+q}|e^{i\mathbf{q\cdot
r}}|m;\mathbf{k}\rangle|^{2}_{\mathbf{q}=q_{y},\mathbf{k}=k_{y}}
\text{.}
\end{split}
\end{equation}

Only $q_{y}$ and $k_{y}$ components are under the numerical calculations, and the evaluated results
remain the same along the other direction of $q_{x}$ and $k_{x}$. Since all the $\pi$-electronic states
are covered in the magneto-electronic excitations the strength and frequency of the resonances in
Im$(-1/\epsilon)$ could be correctly defined. Moreover, effects due to temperatures and high dopings are
allowed.

The magneto-electronic excitation spectra are expected to be greatly diversified by the stacking
configurations in layered graphene systems, as discussed later in Chap.10. The significant effects, which
arise from the magnetic field, the interlayer/intralayer hopping integrals, and the interlayer/intralayer
Coulomb interactions, could be taken into account simultaneously by using the layer-dependent RPA in
Eqs.(4)-(7). The main reason is that the generalized/magnetic tight-binding model is consistent with the
modified RPA under the concept of layer projection. That is to say, the LL energy spectra and wave
functions in few-layer systems are first evaluated from the generalized tight-binding model. The
layer-dependent response functions, which determine the single-particle magneto-excitations, are
investigated by the layer projections of the LL wave functions. And then, the sensitive dependence of
energy loss spectra on the magnetic field strength could be explored in detail. The similar method is
suitable for other emergent layered materials, such as, silicene and germanene.

\subsection{Electron Energy Loss Spectroscopy $\&$ inelastic x-ray scatterings}

EELS\cite{Nature468;1088,Carbon36;561,PRB59;5832,PRB61;5751,JCP114;7477,CPL305;225,
CPL324;255,MSEB74;206,PRB88;075433,Carbon114;70,PRB91;045418,ACSNano12;1837,PRB84;155416,
Nature448;57,MatSci74;351,NanoLett15;4973,Nature487;82,PRB77;233406,APL99;082110,PRB80;113410,
NanoLett14;3827,PRB83;161403,APL102;111609,Plasmonics7;369,Carbon50;183,PRB90;125125,PRB31;4773,
APL77;238,PRB28;3439,PRL42;666,PRL89;076402,SurSci454;462,PSSB81;227,Ultramicroscopy106;1091,
Micron34;235,JMicrosc194;203,RSI63;2195,HIbach,Ultramicroscopy107;575,Egerton,ZPhys243;229,
OptCom1;119,JESRP195;85,RPP72;016502,RMP82;209,APL106;203102,PRL100;196803,PRB95;195411,
SurSci602;2069,Carbon37;733,PRB64;195404,PRB64;115424,PRL80;4729,JJAP31;L1484,PRB49;2882,
APL89;213106,SurSci601;L109,NJP14;103045,JPCM25;345303,SurSci608;88,EPL97;57005,
JPCM23;112204,PRB78;201403,PRB84;033401} and
IXS\cite{Winfried,PRB38;2112,RSI82;113108,RSI77;053102,RMP73;203,RMP79;175,RMP83;705,PRB86;245430,
PRB76;035439,PRB89;014206,JPCM19;046207,JPSJ84;084701,PRB71;060504,PRB90;125125,PRL101;266406}
are the only two kinds of very efficient methods in examining/verifying excitations spectra in
condensed-matter systems. They have been successfully utilized to identify the Coulomb excitations and
phonon dispersion spectra of any dimensional (0D-3D) materials. There exist both inelastic transmission
and reflection EELS, in which the latter is suitable for thoroughly exploring the low-energy excitations
lower than ${1}$eV.
\cite{SurSci602;2069,APL89;213106,SurSci601;L109,NJP14;103045,JPCM25;345303,APL99;082110,
APL102;111609,SurSci608;88,EPL97;57005,JPCM23;112204,PRB78;201403,PRB83;161403,PRB84;033401}
A long-time development for EELS is done since the first measurement on bulk graphite in
1969.\cite{OptCom1;119,ZPhys243;229} The resolutions of transferred energies and momenta are under the
current investigations. IXS just starts to experience a rapid growth within the recent decades, so this
technique need to be greatly enhanced by the various manners. As a result, most of electronic excitations
in carbon/graphene-related systems are accurately measured by EELS. These two techniques have their own
advantages, being reliable in the different environments, as discussed later.

The REELS instrumentation extracts the bulk and surface energy loss functions of the back-scattered
electrons from the sample surface.\cite{Egerton}
If the incident electron beams have a kinetic energy with a few hundred eV, the relatively simple
technique can provide loss spectra with an energy resolution of a few meV, which is sufficient to resolve
vibrational and electronic excitation modes.\cite{Ultramicroscopy107;575}
REELS is widely used for investigating the physical and chemistry properties of material
surfaces.\cite{HIbach}
It typically operates with 25 meV energy resolution in the energy range between 15 and 70
eV,\cite{RSI63;2195} while controlling the momentum resolution down to 0.013 ${\AA}^{-1}$ (better than
one percent of a typical Brillouin zone).\cite{Kittel}
However, the energy resolution can be possibly close to 1 meV under the condition where much weaker
electron beams are adopted using a high-resolution monochromator at an ultrahigh vacuum base pressure
($\sim2\times10^{-10}$ Torr).\cite{HIbach}
On the other hand, in transmission EELS instruments, the incident electron beams pass through the sample
and can be adopted for use in transmission electron microscopy (TEM) to detect the material structure.
This technique commonly understood as EELS.
The spectroscopy uses higher energy electron beams, typically 100-300 keV, as employed in the TEM. The
energy loss is appreciable, typically varying from a few eV up to hundreds eV, with an energy resolution
$<$1 eV and momentum resolution ~0.01 ${\AA}^{-1}$.
In particular, the improvements for electron monochromators and spectrometers make the achievement of an
energy resolution $<$50 meV.\cite{Ultramicroscopy96;367,Micron34;235}
The dedicated TEELS instruments have excellent momentum resolution, and at best 80 meV energy resolution;
they are suitable for measuring the electronic excitations down to 0.5 eV under the consideration of the
interference from the zero-loss line.\cite{JMicrosc194;203,Ultramicroscopy106;1091} The spectral
resolution should be sufficient for studying collective excitations in most metallic and doping
semiconductor materials.
The dispersion relations of the plasmons can be measured by an angle-resolved
EELS, which is performed with low energy electrons and uses an analyzer to detect the scattered
electrons.
The analyzer is a magnetic-prism system, as shown in Fig. 2-2, where the commercially available Gatan
spectrometer is installed beneath the camera and the basic interface and ray paths are shown as well.
The surface of the prism is curved to reduce the spherical and chromatic aberrations.
Scattered electrons in the drift tube are deflected by the magnetic field into a variable entrance
aperture (typically variable from 1 to 5mm in diameter).
All the electrons in any direction are focused on the dispersion plane of the spectrometer; electrons
that lose more energy deflect further away from the zero-energy loss electrons according to the Lorenz
force law.
The magnetic prism projects energy-loss spectrum of the electrons onto a CCD camera, which is
straightforward to capture the whole energy distribution simultaneously.
It is possible to modulate the resolution of the transfered momentum by varying the half-angle of the
incident beam in TEM and the scattering beam in the spectrometer.
The momentum resolution is typically in the order of $\sim0.01{\AA}^{-1}$, based on the angle variation
of a few milliradians.



IXS can directly probes the microscopic dynamic behavior in nanoscale systems.
It has been successfully utilized to detect a wide range of physical phenemena, such as phonon dispersion
in solids, dynamics of disordered materials and biological systems, as well as electronic excitations in
condensed matter systems.\cite{Winfried}
The transferred energy and momentum are independent variables and cover the full spectrum of the
dielectric response.
The medium inelastic x ray beam line is designed to provide high photon flux over the typical Brillouin
zone sizes; the photon energy is distributed from 4.9 keV to 15 keV, with an energy resolution of
$\sim$70 meV and momentum resolution of $\sim0.02-0.03{\AA}^{-1}$.
In particular, the instrument in the Swiss Light Source has extremely good energy resolution of 30
meV.\cite{JSR17;631}
The experimental resolution is possible achieve to a few meV with the development of new synchrotron
sources.
Furthermore, IXS can be used to measure all kinds of electronic excitations
because the electronic charges can interact with X-rays.
Using hard X-ray synchrotron sources, the spectroscopy is a powerful technique to detect the interior
properties of bulk materials, and also can be applied to the systems under external electric and magnetic
fields.\cite{RSI88;033106}
Depicted in Fig. 2-3, the analyzer, built on the basis of Bragg optics, efficiently collect and analyze
the energies and momenta of the scattered photons in a small space, and provides detailed information on
the intrinsic electronic properties of the system.
In order to maximize the scattered photon intensity, a spherically bent analyzer (typically 10 cm in
diameter) is used to capture the scattered radiation of the momentum-transfer photons in a small solid
angle.
The transferred energy is projected onto a CCD detector and the full energy-loss spectrum is scanned by
varying the Bragg angle of the crystal.
Operated in the Rowland circle geometry, the measured double differential scattering cross-section
describes the elementary excitations in the characteristic energy-loss regime via the
dissipation-fluctuation theorem.

There are important differences between EELS and IXS techniques in measuring environments. The incident
particle beam could be focused into $\sim$10 $\AA$ and 100 $\AA$ for the former and the latter,
respectively.\cite{RSI63;2195,Ultramicroscopy107;575,Egerton,HIbach} Furthermore, EELS has more excellent
resolutions in transferred energies and momenta, compared with IXS. Much more inelastic scattering events
could be measured using EELS within a short time; that is, the EELS measurers for the complete excitation
spectra would be done more quickly and accurately. EELS is very suitable for the low-dimensional systems
and nanoscaled structures, since the electron beam simultaneously provides the information of material
size and  position. However, IXS from the continuous synchrotron radiation exhibits a very strong
intensity with the tunable energy and momentum. The extreme surrounding environments, accompanied with
the applications of magnetic/electric fields and different temperatures/pressures, could be overcome
under the inelastic light scatterings. The external fields strongly affect the incident charges and the
sample chamber is too narrow, so that EELS can not work under such environments. For example, the IXS
measurements are very useful in examining/verifying the inter-LL excitations and magnetoplasmons in
graphene-related systems.

The high-resolution EELS could serve as a very powerful experimental technique to explore the Coulomb
excitations in carbon-related systems and emergent 2D materials. The experimental measurements have
successfully confirmed certain electronic excitations due to the $\pi$ and $\sigma$ carriers in the
${sp^2}$ bonding systems, such as,
graphites,\cite{ZPhys243;229,PR138;A197,PR178;1340,PRB7;2275,PRL89;076402,OptCom1;119,SurSci454;462}
graphite intercalation compounds,\cite{PRB31;4773,APL77;238,PRB28;3439,PRL42;666} single- and
multi-walled carbon
nanotubes,\cite{Carbon37;733,PRB64;195404,PRB64;115424,PRL80;4729,JJAP31;L1484,PRB49;2882} single- and
few-layer
graphenes,\cite{PRB88;075433,Carbon114;70,PRB91;045418,ACSNano12;1837,PRB84;155416,Nature448;57,
MatSci74;351,NanoLett15;4973,Nature487;82,PRB77;233406,APL99;082110,PRB80;113410,
NanoLett14;3827,PRB83;161403,APL102;111609,Plasmonics7;369,Carbon50;183,NJP14;103045,JPCM25;345303,
SurSci608;88,EPL97;57005,JPCM23;112204,PRB78;201403,PRB84;033401} $C_{60}$-related
fullerenes,\cite{CPL324;255,MSEB74;206,PRB61;5751} carbon onions,\cite{Carbon36;561,PRB59;5832,
PRB61;5751,JCP114;7477,CPL305;225} and graphene nanoribbons.\cite{Nature468;1088} In general, the
interlayer interactions, dimensions, geometric symmetries, stacking configurations, and chemical dopings
might induce many/some/few free conduction electrons/valence holes, leading to the low-frequency acoustic
or optical plasmon modes with frequency about ${\omega_p\,<1}$ eV. For example, under room temperature,
the 3D Bernal graphite possess the low-frequency optical plasmons at 45$-$50 meV and 128 meV under the
long wavelength limit, respectively, corresponding to the electric polarizations parallel and
perpendicular to the $z$ axis.\cite{JPSJ69;3781} Specifically, the detailed temperature-dependent EELS
show that the former are very sensitive to temperature effects.\cite{JPSJ69;3781} The high-density free
electrons and holes, which are, respectively, created in the donor- and accepter- type graphite
intercalation compounds (layered superlattice systems), possess the ${\sim\,1}$-eV optical plasmons due
to the coupling of collective excitations on infinite layers, in which they strongly depend on the
transferred momenta.\cite{PRB55;13961} On the other hand, for the layered graphene systems with the
adatom chemisorptions,\cite{CRCSYLin} the low-frequency plasmons become 2D acoustic modes; that is, they
have the $\sqrt q$ dependence at small transferred moment.\cite{PRB86;195424}

The $\pi$ plasmons, which are due to all the valence $\pi$ electrons, are found to exist in
carbon-related systems except for diamond. Their frequencies are higher than 5 eV and have a very strong
momentum
dependence.\cite{OptCom1;119,ZPhys243;229,SurSci454;462,PRL42;666,PRB28;3439,APL77;238,PRB31;4773,
PRL100;196803,PRB95;195411,Carbon37;733,Carbon114;70,PRB91;045418,APL99;082110,NanoLett14;3827,
PRB83;161403,Plasmonics7;369,PRB84;155416} For example, the $\pi$-plasmon frequencies in Bernal graphite
grow from ${\sim\,7}$ eV to 11 eV as the transferred momentum in the range of ${0<q<1.4}$
${\AA^{-1}}$.\cite{OptCom1;119,ZPhys243;229,SurSci454;462} The similar $\pi$-plasmom modes are revealed
in layered graphenes and carbon nanotubes.\cite{PRL42;666,PRB28;3439,APL77;238,PRB31;4773,
PRL100;196803,PRB95;195411,Carbon37;733,Carbon114;70,PRB91;045418,APL99;082110,NanoLett14;3827,
PRB83;161403,Plasmonics7;369,PRB84;155416} Their frequencies are enhanced by the increase of graphene
layers.\cite{PRB84;155416} Moreover, there exist several discrete interband plasmon modes in cylindrical
nanotube systems,\cite{PRL80;4729} in which their momentum dependences are very weak.  They belong to
optical modes and possess the frequencies of 0.85 eV. 1.25 eV, 2.0 eV, 2.55 eV; 3.7 eV, indicating the
inter-$\pi$-band excitation mechanisms closely related to the low-lying occupied valence and unoccupied
conduction energy subbands. It should be noticed that such plasmons are in sharp contrast with the
free-carrier-induced acoustic modes (discussed earlier).
As to the ${\pi\,+\sigma}$ plasmons, they arise from the collective excitations of the ${\pi\,+\sigma}$
occupied electrons. Apparently, the super high carrier density will create the pronounced fluctuations
accompanied with very large resonance frequencies.
In general, the ${\pi\,+\sigma}$ plasmon frequencies are higher than 20 eV and strongly dependent on the
transferred momenta.\cite{NanoLett14;3827}
However, they might appear at lower frequency of ${\sim\,15}$ eV, e.g., those in few-layer graphenes.
That all the $\pi$ electrons and part of $\sigma$ ones take part in such collective excitations could
explain this result. In short, the above-mentioned four kinds of plasmon modes are examined/identified by
the high-resolution EELS, while the inelastic light scattering measurements are absent up to now. The
latter are very suitable for the direct verifications of the $\pi$ and ${\pi\,+\sigma}$ plasmons with
higher frequencies.

\subsection{Coulomb decay rates \& ARPES}

The many-body self-energy  is derived to characterize the valence quasi-particle properties in layered
graphene-related systems with valence and conduction bands. It is suitable under any temperature and
doping. Furthermore, the relation between the quasi-particle energy widths and the ARPES measurements is
discussed in detail. In addition, three kinds of femtosecond pump-probe spectroscopies are useful in
comprehending the lifetimes/energy widths of the specific electronic states are also discussed.

\subsubsection{Coulomb decay rates in layered graphene-related systems}

The incident electron beam and electromagnetic field, which act on the layered graphene-related systems,
will be dynamically screened by conduction and valence electrons, or will have strong interactions
between the external perturbations and the charge carriers. During the complicated screening processes,
they create the excited electrons (holes) above (below) the Fermi level. Such intermediate states could
further decay by the inelastic electron-electron and electron-phonon scatterings. At low temperature, we
only focus on the former mechanisms. The Coulomb decay rate (${1/\tau}$) is fully determined by the
effective interaction potential (${V^{eff}}$) between two charge carriers, in which the dynamic e-e
interactions could be understood from the layer-dependent modified RPA. By using the Matsubara Green's
functions,\cite{Mahan} ${1/\tau}$ of monolayer graphene is calculated from the quasiparticle self-energy,
the screened exchange energy (the RPA self-energy as clearly shown in Fig. 2-4)

\begin{equation}
\Sigma(\mathbf{k},h,ik_{n})=-\frac{1}{\beta}\sum_{\mathbf{q},h^{\prime},i\omega_{m}}V^{eff}
(\mathbf{k},h,h^{\prime};\mathbf{q},i\omega_{m})G^{(0)}(\mathbf{k+q},h^{\prime},ik_{n}+i\omega_{m})\text{,}
\end{equation}

where $\beta=(k_{B}T)^{-1}$, $ik_{n}=i(2n+1)/\pi/\beta$ (complex fermion frequency),
$i\omega_{m}=i2m\pi/\beta$ (complex boson frequency) and $G^{(0)}$ is the noninteracting Matsubara
Green's function.
$V^{eff}(\mathbf{k},h,h^{\prime};\mathbf{q},i\omega_{m})=V(\mathbf{k,q},h,h^\prime\,)/\epsilon\,({\bf
q},i\omega_m\,)=V_{q}|\langle h^{\prime},\mathbf{k+q}|e^{i\vec{q}\cdot
\vec{r}}|h,\mathbf{k}\rangle|^{2}/[\epsilon(\mathbf{q},i\omega_{n})]$ is the screened Coulomb
interactions with the band-structure effect, in which the intraband and the interband deexcitation
channels need to be taken into account simultaneously. $V_{q}$ is the 2D bare Coulomb potential energy
and $\epsilon(\mathbf{q},i\omega_{n})$ is the RPA dielectric function.
This equation is also suitable for monolayer silicene and germanene under the spin-degenerate
states,\cite{PRB97;195302} although they possess the significant spin-orbital couplings. It does not need
to solve the spin-up- and spin-down-related Coulomb decay rates separately, since they make the same
contributions. That is, it is sufficient in fully exploring the wave-vector-, conduction/valence- and
energy-dependent self-energies (Eq. 14).

Under the analytic continuation $ik_{n}\rightarrow E^{h}(\mathbf{k})$, the carrier self-energy could be
divided the bare exchange energy, line part and residue part:

\begin{equation}
\Sigma_{sx}(\mathbf{k},h,E^{h}(\mathbf{k}))=\Sigma_{x}({\bf k},h,E^h({\bf
k}))+\Sigma^{(line)}(\mathbf{k},h,E^{h}(\mathbf{k}))+
\Sigma^{(res)}(\mathbf{k},h,E^{h}(\mathbf{k}))
\text{,}
\end{equation}

in which

\begin{equation}
{\Sigma_{x}\,({\bf k},h,E^h({\bf k}))=-\sum_{{\bf
q},h^\prime}V(\mathbf{k,q},h,h^\prime\,)f(E^{h^\prime}({\bf k+q})}),
\end{equation}

\begin{equation}
\begin{split}
\Sigma^{(line)}(\mathbf{k},h,E^{h}(\mathbf{k}))
&=-\frac{1}{\beta}\sum_{\mathbf{q},h^{\prime},i\omega_{m}}
[V^{eff}(\mathbf{k},h,h^{\prime};\mathbf{q},i\omega_{m})-V(\mathbf{k},h,h^\prime\,\mathbf{q})]\\
&\times G^{(0)}(\mathbf{k+q},h^{\prime},E^{h}(\mathbf{k})+i\omega_{m})
\text{,}
\end{split}
\end{equation}

and

\begin{equation}
\begin{split}
\Sigma^{(res)}(\mathbf{k},h,E^{h}(\mathbf{k}))=&-\frac{1}{\beta}\sum_{\mathbf{q},h^{\prime},i\omega_{m}}
[V^{eff}(\mathbf{k},h,h^{\prime};\mathbf{q},i\omega_{m})-V(\mathbf{k},h,h^\prime\,\mathbf{q})]\\
&\times[G^{(0)}(\mathbf{k+q},h^{\prime},ik_{n}+i\omega_{m})
-G^{(0)}(\mathbf{k+q},h^{\prime},E^{h}(\mathbf{k})+i\omega_{m})]
\text{.}
\end{split}
\end{equation}

The summation of the line and residue parts is the so-called correlation self-energy. The imaginary part
of the residue self-energy determines the Coulomb decay rate, being
characterized as

\begin{equation}
\begin{split}
Im\Sigma^{(res)}(\mathbf{k},h,E^{h}(\mathbf{k}))=&\frac{-1}{2\tau(\mathbf{k},h)}\\
&=\sum_{\mathbf{q},h^{\prime}}Im[-V^{eff}(\mathbf{k},h,h^{\prime};\mathbf{q},\omega_{de})]\\
&\times\{n_{B}(-\omega_{de})[1-n_{F}(E^{h^{\prime}}(\mathbf{k+q}))]
-[n_{F}(E^{h^{\prime}}(\mathbf{k+q}))]\}\\
&=\frac{-1}{2\tau_{e}(\mathbf{k},h)}+\frac{-1}{2\tau_{h}(\mathbf{k},h)}
\text{.}
\end{split}
\end{equation}

$\omega_{de}=E^{h}(\mathbf{k})-E^{h^{\prime}}(\mathbf{k+q})$ is the deexcitation/decay energy. $n_{B}$
and $n_{F}$ are, respectively, the Bose-Einstein and Fermi-Dirac distribution functions. Equation (17)
clearly means that an initial state of $(\mathbf{k}; h)$ can be deexcited to all the available
$(\mathbf{k+q}; h^{\prime})$ states under the Pauli exclusion
principle and the conservations of energy and momentum. The excited states above or below
the Fermi level respectively, exhibit the electron and hole decay rates (the first and
second terms in Eq. (20)). By the detailed derivations, the zero-temperature Coulomb decay
rates of the excited electrons and holes are

\begin{equation}
\begin{split}
\frac{1}{\tau_{e}(\mathbf{k},h)}+\frac{1}{\tau_{h}(\mathbf{k},h)}=&-2\sum_{\mathbf{q},h^{\prime}}
Im[-V^{eff}(\mathbf{k},h,h^{\prime};\mathbf{q},\omega_{de})]\\
&\times[-\Theta(\omega_{de})\Theta(E^{h^{\prime}}(\mathbf{k+q})-E_{F}))+[\Theta(-\omega_{de})
\Theta(E_{F}-E^{h^{\prime}}(\mathbf{k+q}))]
\text{.}
\end{split}
\end{equation}

where $E_{F}$ is the Fermi energy for a pristine system/an extrinsic system with carrier doping. $\Theta$
is the step function that limits the available deexcitation channels. The Coulomb decay rate is double
the energy width of a quasi-particle state. Equations (19) and (20) could be generalized to a
single-walled carbon nanotube with a cylindrical symmetry.\cite{PhysicaE34;658}

The layer-projection method could be developed to thoroughly investigate the Coulomb decay rates in
few-layer graphene-related systems. Any electronic states are composed of the tight-binding functions
localized at the different layers, so that their inelastic Coulomb scatterings are closely related to the
effective layer-dependent Coulomb potentials (${V^{eff}_{ll^{\prime}}}$’s). First, we need to evaluate
${V^{eff}_{ll^{\prime}}}$ in Eq. (4) by using the analytic and numerical forms simultaneously. And then,
Eq. (20) is directly suitable for studying the decay rates. The various deexcitation channels are similar
in the layer-dependent ${V^{eff}_{ll^{\prime}}}$, but they might exhibit the distinct weights. However,
the calculations become very heavy even under the tight-biding model. On the other hand, there are only
few studies on bilayer graphenes up to now,\cite{PLA357;401} in which the main decay mechanisms are not
clear in the first-principles method because of the numerical resolution.

\subsubsection{ARPES measurements on occupied quasi-particle energy widths}

ARPES is the most efficient $\&$ reliable equipment in studying the quasi-particle band dispersions and
energy widths for the occupied electronic states within the first Brillouin zone. Their measurements
could examine the band-structure calculations by the tight-bindung model and the first-principles method,
and the predicted Coulomb decay rates under the screened exchange self-energy. In general, the ARPES
chamber is combined with the instruments of sample synthesis to measure the in-situ quasi-particle
states. When a specific condensed-matter system is illuminated by the soft X-ray (Fig. 2-5), the occupied
valence states are excited to the unoccupied intermediate ones under a electric-dipole perturbation.
Photoelectrons are excited by incident photons and escape outside of the material surface into the
vacuum, and then they are measured by an angle-resolved (energy,momentum) analyzer. The total momenta of
photoelectrons are evaluated from the electron gas model, in which the parallel and perpendicular
components depend on the polar and azimuthal angles, as shown in Fig. 2-5 by $\theta$ and $\varphi$,
respectively. The former is conserved through the photoemission process, while the conservation law is
not reliable for the latter because of the destruction of translation symmetry along the direction normal
to surface. As a result, ARPES measurements are mainly focused on two- and quasi-two-dimensional systems
with the negligible energy dispersions perpendicular to surface.
However, the non-conservation issue might be solved using the important characteristics of the
$k_z$-dependent band structure, as done for the 3D band structure of layered graphite by that at
${k_z=0}$.\cite{PRL100;037601,NatPhys2;595,ASS354;229,PRB79;125438} Specifically, the ARPES measurements
could provide energy widths of valence states, directly reflecting the many-particle deexcitation
scatterings arising from the electron-electron and electron-phonon interactions. Improvements in energy
and momentum resolutions have become a critical factor for studying the emergent low-dimensional
materials. Up to date, the best resolutions for energy and angular distribution are, separately, $\sim$1
meV and 0.1$^{\circ}$ in the UV region.

The high-resolution ARPES is an only experimental instrument in directly measuring the
wave-vector-dependent valence/occupied energy spectra. The experimental measurements have confirmed the
geometry-enriched band structures in the graphene-related condensed-matter systems, as verified for
various dimensions, layer numbers, stacking symmetries, and adatom/molecule chemisorptions. There exist
${(k_x,k_y,k_z)}$-dependent 3D band structures of Bernal
graphite,\cite{PRL100;037601,NatPhys2;595,ASS354;229,PRB79;125438}
1D parabolic energy subbands in graphene nanoribbons,\cite{ACSNano6;6930,PRB73;045124}
the linearly isotropic Dirac-cone structure in monolayer graphene,\cite{PRL110;146802,NatPhys3;36} and
few-layer AA-stacked graphene,\cite{NatMater23;887,NanoLett17;1564}
two pairs of parabolic dispersions in AB-stacked bilayer
graphene,\cite{PRL98;206802,PRB88;075406,PRB88;155439,NPGAsiaMat10;466}
the coexistent linear and parabolic bands in symmetry-broken bilayer graphene,\cite{NatMater12;887} the
linear and parabolic bands in trilayer graphene with ABA
stacking,\cite{PRL98;206802,PRB88;155439,NPGAsiaMat10;466}
the linear, partially flat and sombrero-shaped bands in ABC-stacked trilayer
graphene,\cite{PRB88;155439,PRL98;206802,NPGAsiaMat10;466}
the metal-semiconductor transitions and the tunable low-lying energy bands after the molecule/adatom
absorptions on graphene surface.\cite{ACSNano6;199}
On the other side, the predicted Coulomb decay rates could be examined from the high-resolution ARPES
measurements on the energy widths, as clearly revealed in potassium chemiadsorption on monolayer
graphene. The ARPES energy spectra are done along KM and K$\Gamma$ directions under various doping
concentrations of monolayer electron-doped graphene, obviously indicating the linewidth variation with
wave vector. They are further utilized to analyze  the doping-dependent momentum distribution curves
(MDCs). The Lorentzian peak forms are centered at the quasi-particle energies; furthermore, they exhibit
the full width at the half-maximum intensity identified as ${-2Im\Sigma^{(res)}}$ (just the scattering
rate). The single-particle excitation and plasmon modes, as well as the electron-phonon scatterings at
finite temperatures, are proposed to comprehend the unusual energy dependences of the MDCs linewidths.
The ARPES measurements at low temperatures could provide the Coulomb-scattering-dominated MDCs to verify
the theoretical calculations.

In addition to the direct ARPES measurements on the energy widths of the valence quasiparticle states,
the lifetimes (the inverse of the former) of the specific states, including the Fermi-momentum states and
the excited valence and conduction band-edge states, could be examined by three kinds of pump-probe [also
sees Chap. 14.1]. The femtosecond photoelectron spectroscopy is available in fully exploring the carrier
relaxation near the Fermi level; that is, it is very suitable for the semimetallic and metallic systems.
For example, the measured liftimes, which correspond to the Fermi-momentum states in the Bernal graphite
and the metallic single-walled and multiwall carbon nanotubes. are, respectively, ${\tau\sim\,0.5}$ ps
and ${\tau\sim}$0.2 ps at room temperature.\cite{NatPhys3;36,NatComm5;3257,PRL102;107007,JPSJ73;3479,
PRL84;5002,PRL76;483,PRL87;267402}
As for the femtosecond optical absorption/transmission/reflectivity and fluorescence spectroscopies, they
are designed for the semiconducting systems, such as the type-II narrow-gap and type-III moderate-gap
carbon nanotubes.
The latter nanotube systems are identified to exhibit the lifetimes of ${\tau\,\sim\,0.3-1.5}$ ps and
${\tau\,\sim\,0.3-1.5}$ ps, respectively, being associated with the first and second prominent absorption
peaks.\cite{PRB42;2842,PRL90;057404,PRL92;017403,PRL92;117402,PRL93;017403,
PNASU115;674,PRB42;2842,PRB97;075435,Science288;492}
Such decay rates are attributed to the intraband inelastic Coulomb scatterings
The experimental measurements on carbon nanotubes are consistent with the theoretical
prdictions.\cite{PhysicaE34;658} The time- and temperature-dependent photolumence spectra have been made
on the very small ${(6,4)}$ carbon nanotube in the range of ${48-182}$ K, revealing the band-edge-state
lifetimes due to the first pari of energy bands about ${100-20}$ ps.\cite{PRL95;197401,PRL92;17740} Such
femtosecond spectroscopies are the critical tools in studying the generation, relaxation, and
recombination of the nonequilibrium charge carriers, i.e., they can probe and verify the time-dependent
carrier dynamics.

\section{ Concluding remarks and perspectives}

The current book clearly presents a fully modified theory on Coulomb excitations/decays in
graphene-related systems, in which the theoretical framework combines the layer-dependent RPA and the
generalized tight-binding model. It can deal with a plenty of critical factors related to the different
lattice symmetries, layer numbers, dimensions, stacking configurations, orbital hybridizations,
intralayer $\&$ interlayer hopping integrals, spin-orbital couplings, temperatures, electron/hole
dopings, electric field, and magnetic field.
Apparently, there exist the rich and unique electronic excitation phenomena due to the distinct energy
bands and wave functions in the various condensed-matter systems, as obviously revealed in the diverse
(momentum, frequency)-phase diagrams.
The calculated results, with the concise physical pictures, clearly illustrate the very important roles
of the e-e Coulomb interactions.
Of course, they could explain the up-to-date experimental measurements. This model could be generalized
to the other emergent 2D materials under the detailed calculations/investigations, such as, the layered
silicene,\cite{PRB94;205427,RSCAdv5;51912,PRB97;125416,NJP16;125002}
germanene,\cite{SciRep7;40600,PRB97;195302} tinene,\cite{SciRep7;1849} phosphorene,\cite{IOPChen}
antimonene,\cite{IOPChen} bismuthene,\cite{IOPChen,NJP20;062001} and MoS$_2$.\cite{IOPChen} The further
studies would provide the significant differences among these systems and be very useful in thoroughly
understanding the close/complicated relations of the essential physical properties.
On the other hands, the theoretical models should be derived again to solve the Coulomb excitations in 1D
and 0D systems without the good spatial translation
symmetry.\cite{PRB76;115422,PRB47;6617,PRB76;115422,JPSJ68;3806,PRB53;15493,PRB56;1430,PRB57;10183} For
example, 1D graphene nanoribbons and 0D graphene quantum dots have the open boundary conditions, so that
they, respectively, possess many energy subbands and discrete energy levels. Maybe, the dielectric
function tensor, being characterized by the subband/level index, is one effective way to see the
excitation properties.\cite{PCCP18;7573}

The theoretical framework of Coulomb excitations and decay rates have been fully developed in Chap. 2, in
which the experimental progresses, respectively, on EELS, IXS and ARPES are investigated in detail. The
dielectric functions and the energy loss functions are, respectively, are responsible for the single- and
many-particle excitation spectra, They strongly depend on the geometric structure/the translation
symmetry, directly affecting the longitudinal transferred momenta. As a result, the 3D graphite, 2D
graphene, and 1D carbon nanotubes, respectively, exhibit the dimension-related bare Coulomb interactions.
so that such condensed-mattered systems are expected to create the diverse   Coulomb excitation
phenomena. Specifically, few-layer graphenes possess the tensor forms in the dielectric functions, but
not the scalar quantities. The main reason is that the interlayer hopping integrals and the interlayer
Coulomb interactions need to be taken into account simultaneously. The straightforward combinations
between the modified RPA and the generalized tight-binding model has been made; furthermore, the
dimensionless energy loss function, directly corresponding to the measured excitation spectrum, is
well-defined and rather reliable. The new theoretical model is very suitable for the composite effects
due to the stacking configurations, the number of layers, the various lattice symmetries, the
spin-orbital couplings, the electric field, and the magnetic field, especially for the unusual
magneto-electronic excitations arising from the magnetic quantization. Also, for the excited
quasiparticles in layered graphenes, it is rather reliable in exploring the Coulomb inelastic scattering
rates of the excited quasiparticles in layered graphenes by linking the screened exchange energy of the
Matsubara$^{,}$s Green functions.  This method could provide the clear and concise physical pictures
about the effective deexcitation channels. On the other side, the experimental techniques, the up-to-date
resolutions, and the whole measured results are another focuses of diverse excitation phenomena. The
detailed comparisons with the theoretical predictions could be found in the following chapters.

Monolayer graphene exhibits the rich and unique Coulomb excitations in the presence/absence of
temperature, doping, and interlayer Coulomb coupling. A pristine system is a zero-gap semiconductor, so
that the Landau dampings at ${T=0}$ only coming from the inter-$\pi$-band transitions of the linear
Dirac-cone structure are too strong to observe the 2D acoustic plasmon modes. The thermal excitations
could create the $T^2$-dependent free electron/hole density. When temperature is higher than the critical
one, the intra-$\pi$-band excitations and the 2D-like plasmons ($\sim$ 0.1 eV at room temperature) come
to exist. Compared with the thermal effects, the carrier doping induces the new/extra single-particle and
collective excitations more efficiently.  Apparently, the higher free carrier density is responsible for
the very prominent asymmetric peaks in the polarization functions and the energy loss ones. The rich
(momentum, frequency)-phase diagram covers the intraband e-h excitations, the interband ones, the vacuum
regions without any excitations, and the undamped/dampled acoustic plasmon mode, being sensitive to the
variation of the Fermi level/the doping density, However, it hardly depends on the electron or hole
doping as a result of the almost symmetric Dirac-cone energy spectrum about ${E_F=0}$. In addition to the
tight-binding model, the effective-mass approximation is utilized to obtain the analytic formula for the
polarization/dielectric function, where the conservation of particle numbers needs to be carefully solved
during the detailed derivation. It should be noticed that the significant differences between a doped
monolayer graphene and a 2D electron gas, directly reflecting the conduction $\&$ valence Dirac cones and
the parabolic conduction band, include  the distinct boundaries of the intraband excitations, the
existence/absence of the interband excitations, the different  Landau dampings of the 2D plasmon modes,
the stronger momentum-dependence  of plasmon frequencies and  more rich (momentum, frequency)-phase
diagram in the former. Up to now, the 2D plasmon modes are identified in the EELS measurements of
alkali-doped graphene
systems,\cite{PRB28;3439,ACSNano12;1837,PRB84;155416,MatSci74;351,NanoLett15;4973,APL99;082110} while the
temperature-induced ones require the further experimental examinations. The Dirac-cone band structure is
approximately reliable in a  double-layer system with a sufficiently long interlayer distance. The
in-phase and out-of-phase collective oscillations, which, respectively, arise from the symmetric and
anti-symmetric superpositions of free carriers on the first and second layers, appear in the energy loss
spectra. They belong to the optical and acoustic plasmon modes according to the momentum dependences. The
superlattice mode, with a Dirac-cone  band structure and the significant interlayer Coulomb interactions,
are very useful in understanding the excitation and deexcitation spectra in graphite intercalation
compounds.\cite{PRB55;13961,PRB34;979}

According to the tight-binding model and the first-principles
method,\cite{CRCPress;9781138571068,CRCPress;9781138556522} the trilayer ABA stacking presents the
composite energy bands, being the superposition of the monolayer- and bilayer-like ones. The
layer-dependent polarization functions are a ${3\times\,3}$  Hermitian matrix, in which the there exist
four independent components. A pristine system only exhibits the obvious interband e-h excitations and
cannot create any plasmon modes, mainly owing to a very low free carrier density and a non-prominent
density of states from the first pair of energy bands. The strong effects, which are due to the electron
doping, can dramatically alter the boundaries of the single-particle excitations, add the new excitation
channels, and induce three kinds of plasmon modes. The first acoustic mode, the second and the third
optical ones, respectively, originate from all the intraband ${\pi_i^c\rightarrow\,\pi_i^c}$ excitations,
the ${\pi_2^c\rightarrow\,\pi_3^c}$ and ${\pi_1^c\rightarrow\,\pi_3^c}$ interband  transitions. The last
kind could be observed only for the higher Fermi level  crossing the highest conduction band. There exist
the diverse ${({\bf q}\,,\omega\,)}$-phase diagrams, being sensitive to the doping carrier density,
stacking configuration and layer number. From the view point of electronic and optical
properties,\cite{CRCPress;9781138571068} the ABA-stacked trilayer  graphene could be regarded as the
superposition of monolayer and bilayer systems under the single-particle schemes. Apparently, it is not
suitable for the many-particle e-e Coulomb interactions. That is to say, it is impossible to understand
the diverse (momentum, frequency)-phase diagrams of a trilayer system from those of the composite ones,
since the different energy bands will have the significant relations during the interband Coulomb
scatterings. Of course, the single-particle and collective excitations are getting more complicated in
the increment of layer number, in which the cross-over behavior between the 2D layered graphene and the
3D Bernal graphite might be worthy of a systematic investigation.

Tri-layer ABC-stacked graphene has three pairs of unusual energy bands near the Fermi level [detials in
Chap. 6.1]; therefore, it exhibits the rich and unique Coulomb excitations. The layer-indexed bare
response functions have nine components, but only four independent ones. There are a lot of
single-particle channels and five kinds of plasmon modes during the variation of free carrier densities.
The latter cover  (I) the intraband plasmon related to the ${\pi_1^c\rightarrow\,\pi_1^c}$ transitions,
(II) the interband mode associated with the ${\pi_1^c\rightarrow\,\pi_2^c}$ excitations, (III) the
interband one arising from the ${\pi_1^v\rightarrow\,\pi_1^c}$ channels, and (IV) $\&$ (V) the multimode
collective excitations under various intraband and interband channels. The complicated relations between
the single-particle and collective excitations create the diverse (momentum, frequency)-excitation phase
diagrams. The plasmon peaks in the energy loss spectra might decline and even disappear under various
Landau dampings. The linear acoustic plasmon is related to the surface states (the partially flat bands
at $E_F$) in pristine system, while it becomes an square-root acoustic mode at any doping. Specially, all
the layer-dependent atomic interactions and Coulomb interactions have been included in polarization
function and dielectric function. The predicted results could be examined from the high-resolution
electron-energy-loss spectroscopy\cite{Nature468;1088,Carbon36;561,PRB59;5832,
PRB61;5751,JCP114;7477,CPL305;225,
CPL324;255,MSEB74;206,PRB61;5751,
PRB88;075433,Carbon114;70,PRB91;045418,ACSNano12;1837,PRB84;155416,Nature448;57,
MatSci74;351,NanoLett15;4973,Nature487;82,PRB77;233406,APL99;082110,PRB80;113410,
NanoLett14;3827,PRB83;161403,APL102;111609,Plasmonics7;369,Carbon50;183,
PRB90;125125,PRB31;4773,APL77;238,PRB28;3439,PRL42;666,PRL89;076402,SurSci454;462,PSSB81;227,Ultramicroscopy106;1091,Micron34;235,JMicrosc194;203,RSI63;2195,HIbach,
Ultramicroscopy107;575,Egerton,ZPhys243;229,OptCom1;119,JESRP195;85,RPP72;016502,RMP82;209,APL106;203102,
PRL100;196803,PRB95;195411,SurSci602;2069,
Carbon37;733,PRB64;195404,PRB64;115424,PRL80;4729,JJAP31;L1484,PRB49;2882,
APL89;213106,SurSci601;L109,NJP14;103045,JPCM25;345303,APL99;082110,APL102;111609,SurSci608;88,EPL97;57005,
JPCM23;112204,PRB78;201403,PRB83;161403,PRB84;033401} and
IXS.\cite{PRB38;2112,RSI82;113108,RSI77;053102,
RMP73;203,RMP79;175,RMP83;705,PRB86;245430,PRB76;035439,PRB89;014206,JPCM19;046207,JPSJ84;084701,
PRB71;060504,PRB90;125125,PRL101;266406} The magneto-electronic Coulomb excitations in ABC-stacked
few-layer graphene systems are expected to the new magnetoplasmon modes because of the
frequent-anti-crossing LLs.\cite{PRB90;205434} Moreover, the theoretical framework of the layer-based RPA
could be further generalized to study the e-e interactions in emergent 2D materials, e.g., ABC-stacked
trilayer silicene and germanene.\cite{PRB94;205427,SciRep7;40600}

Apparently, the trilayer AAB stacking exhibit the unique electronic properties and thus the diverse
Coulomb excitations. The lower stacking symmetry leads to three pairs of unusual energy dispersions: the
oscillatory, sombrero-shaped, and parabolic ones. The former two possess the large and special van Hove
singularities, especially for the first pair nearest to the Fermi level. As a result, for the pristine
system, there exist nine categories of valence$\rightarrow$conduction interband transitions. The special
structures in the bare response functions cover the square-root asymmetric peaks and the shoulder
structures [the pairs of anti-symmetric prominent peaks and logarithmically symmetric peaks] in the
imaginary [real] part. The threshold channel, ${\pi_1^c\rightarrow\,\pi_1^c}$, can create the significant
single-particle excitations and the strong collective excitations. The low-frequency acoustic plasmon,
being characterized by the pronounced peak in the energy loss spectrum, is purely due to the large DOS in
the oscillatory valence and conduction bands and the narrow energy gap; furthermore, its intensity and
frequency is somewhat reduced by the finite temperatures. The similar plasmon mode is revealed in a
narrow-gap carbon nanotube.\cite{JPSJ68;3806} The critical mechanism about the creation of this plasmon
is thoroughly transformed into all the intraband conduction-band excitations, in which the effective
channels and the critical transferred momenta strongly depend on the Fermi level. After the electron/hole
doping, the interband e-h excitation regions are drastically modified and the extra intraband ones are
generated during the variation of $E_F$. Moreover, one or two higher-frequency optical plasmon modes
survive under the various Fermi levels. They are closely related to the specific excitation channels or
the strongly overlapped multi-channels, being sensitive to the Fermi level and transferred momenta. There
are certain important differences among the trilayer AAB, ABC, ABA and AAA stackings, such as, the
boundaries of the various intraband and interband e-h excitations, and the mechanism, number, strength,
frequency and mode of the collective excitations. To fully explore the geometry-enriched Coulomb
excitations, the above-mentioned theoretical predictions require the experimental verifications.

The sliding bilayer graphene systems obviously exhibit the stacking-configuration- and doping-enriched
Coulomb excitation phenomena, mainly owing to the unique essential properties. The shift-dependent
characteristics of electronic properties cover the two distinct Dirac cones with the non-titled/titled
axis, the normal parabolic bands, the highly hybridized/distorted energy dispersions, the stateless
arc-shaped regions, the Dirac points, the local minima/maxima, the saddle points, the Fermi-momentum
states, and the symmetric $\&$ anti-symmetric/the abnormal superposition of the subenvelope functions on
the equivalent/non-equivalent sublattices. The available excitation channels are classified into the
intrapair intraband, interpair, and intrapair interband transitions, strongly depend on the relative
shift of two graphene layers. Under the small transferred momenta, they, respectively, appear at very low
frequency, ${0.2}$ eV${<\omega\,0.4}$ eV, and ${0.7}$ eV${<\omega\,0.9}$. The former two channels are
very sensitive to the free carrier doping; that is, they might come to exist, or be partially/fully
suppressed under the rigid shift of the Fermi level. The  last one hardly depends on $E_F$. The
single-particle excitations, which are associated with the band-edge and Fermi-momentum states, create
the special structures in the polarization functions. The abnormal imaginary/real parts present in the
square-root asymmetric peaks, the shoulders/the logarithmically symmetric peaks, and the inverse
structures of the second types. They, respectively, arise from the Fermi momenta with the linear energy
dispersions, the extremal states of the parabolic bands, and the saddle points. Such prominent
single-particle responses are responsible for the serious Landau dampings. Also, both Fermi momenta and
band-edge states determine the various electron-hole boundaries, where, the latter correspond to the
higher-frequency regions. As to the collective excitations, the acoustic and optical plasmons might exist
simultaneously, only one mode survives, or both of them are absent, strongly relying on the relative
shift and the free carrier density. The single- and many-particle excitations keep the similar behaviors
under a small relative shift of two graphene layers. The [${\delta\,=0}$
${\delta\,=1/8}$]/[${\delta\,=6/8}$, ${\delta\,=1}$]/[${\delta\,=11/8}$, ${\delta\,=12/8}$] bilayer
stackings are similar to each other, while these three sets sharply contrast with one another. The
plasmon modes are most easy to be observed in the first set, and the opposite is true for the third set.

A uniform perpendicular electric field has created the significant effects on the electronic properties
and thus diversifies the Coulomb excitation phenomena. The essential properties, which over the Dirac
points, the Fermi velocity, the Fermi momenta, the valence and conduction band overlap/the free electron
and hole densities, the band gap, the band-edge states with large DOSs, the energy dispersions, the
symmetric $\&$ antisymmetric linear superposition of the tight-binding functions, and the equivalence of
the layer-dependent A and B sublattices, clearly display the drastic changes or the dramatic
transformations under the layer-related  Coulomb potential on-site energies. This field is further
responsible for the unusual single-particle and collective excitations in layered graphene systems.
Apparently, the AA-, AB-, and ABC-stacked few-layer graphenes exhibit the diverse excitation behaviors.
For the $N$-layer AA stackings, there is two/one acoustic plasmon modes in the presence/absence of the
electric field. The higher- and lower-frequency modes, respectively, come from all the free carriers (the
intrapair intraband transitions) and the $F$-induced  band disparity, so that the former behaves like
that in a 2D electron gas, and the latter comes to exist only under the significant splitting of the
intrapair intraband and interband transitions at the enough large $F$$^{,}$s. More optical plasmon modes
is revealed in the energy loss spectrum except for the bilayer stacking, compared with the ${N-1}$ modes
in the pristine system. Such optical plasmons mainly arise from the interpair interband channels;
furthermore, the extra modes are associated with the nonuniform Fermi velocities and energy spacings
between the different Dirac points. Specifically, the AA bilayer stacking can clearly illustrate the
concise physical mechanisms by analytically evaluating the electric-field-dependent band structure, wave
functions, bare polarization functions (intrapair and interpair parts), and dielectric function matrix.
For example, the vanishing determinant of the last one is very useful in understanding the
plasmon-frequency dispersion relations with the transferred momentum. Apparently, the $F$-induced
electronic excitations are very sensitive to the stacking configurations. The significant differences
among the trilayer AAA, ABA and ABC stackings lie in the main features of electron properties, the
available transition channels, the singular structures in bare polarization functions; the e-h excitation
boundaries and the plasmon modes in the ${(q, \omega\,)}$- $\&$ ${(F, \omega\,)}$-phase diagrams. During
the variation of $F$, three pairs of vertical Dirac-cone structures remain similar in the AAA stacking,
while the others present the strongly oscillatory energy dispersions, especially for the first pair of
valence and conduction bands. Most important, the semimetal-semiconductor transition is only revealed
under the specific ABC stacking. The observable single-particle transition channels, which correspond to
the AAA-, ABA- and ABC-stacked systems, are, respectively, divided into three interband and two intraband
excitation categories, five interband excitation categories, and four ones. Furthermore, such bare
polarization functions, being associated with the band-edge states, display the square-root divergent
peaks and the shoulder $\&$ asymmetric peaks (composite structures of shoulders and logarithmically
divergent peaks) in the AAA and [ABA, ABC] stackings. The most prominent plasmon, respectively, behaves
the acoustic and optical modes for the [AAA, ABA] and ABC stackings. The mechanism is different between
AAA and [ABA, ABC] stackings, since this mode mainly originate from the first $\&$ third pairs of Dirac
cones and the first pair of valence and conduction bands. Within a certain range of ${(q, F)}$, there
co-exist  the splitting modes, in which the number is, respectively, two, two and three. As to the
higher-frequency optical plasmons, it is relatively easy to observe them in the trilayer AAA stacking
because of three splitting modes. However, only one mode appears in the ABA and ABC systems.

The entire LL energy spectrum is included in the calculations. This ensures the correctness of the
dielectric function within the RPA and consequently the energy loss function, and the intensity and
frequency of magnetoplasmon modes. The exact diagonalization method in efficiently solving the LL
wavefunctions and the Coulomb matrix elements could also be applicable to multilayer graphene or bulk
graphite over a wide range of magnetic and electric fields. As to monolayer graphene, there are a lot of
observable and weak magneto-plasmon peaks, being sensitive to the strength of the e-h dampings. Their
frequencies present the unusual $q$-dependence, in which the critical transferred momentum directly
reflects the rather strong competition between the longitudinal Coulomb forces and the transverse
cyclotron ones. The positive and negative group velocities, which, respectively, occur at  ${q<q_{B}}$
and ${q>q_{B}}$, corresponds to the dominance of the former and the latter. $q_{B}$ and the
magnetoplasmon frequencies grows with the magnetic field, especially for ${\omega_p\,\propto\sqrt B_z}$.
Temperature effectively induce certain low-lying intraband excitations due to the conduction/valence LLs.
The intraband magnetoplasmons would come to exist only under the sufficiently high temperature, in which
the critical temperature grows with the increasing magnetic-field strength. The magneto-Coulomb
excitations are greatly diversified by the  stacking configurations. The low-energy excitations in the
bilayer AB/AA stacking are only created by the interband/intrabands of the intragroup inter-LL channels.
This directly reflects the spatial symmetric or anti-symmetric distributions on four sublattices of the
magnetic LL wave functions, as observed in zero-field cases. The former and the latter, respectively,
exhibit a lot of and certain prominent peaks in the layer-dependent polarization functions and the energy
loss spectra. The discrete magnetoplasmon frequencies of the AB stacking are similar to those in
monolayer graphene, while they present more complicated momentum-dependence. The critical momenta and the
serious Landau dampings are responsible for the discontinuous but monotonous $B_z$-dependence. Specially,
the 2D-like plasmon, accompanied with some discrete modes, appears in the AA stacking, clearly
illustrating free holes/electrons in the unoccupied valence LLs of the first group/the occupied
conduction ones of the second group. The dramatic variation of the highest occupied LL leads to the
discontinuous and oscillatory $B_z$-dependence. Moreover, the significant differences between the Coulomb
and EM wave perturbations clearly indicate the diverse magenta-electronic excitation phenomena.

Apparently, the 3D three kinds of graphites show the diverse Coulomb excitation phenomena in the low- and
middle-frequency ranges, mainly owing to the dimensionality and stacking configurations. That is, the
single-particle and collective excitation fully reflect the main features of band structures, the strong
wave-vector dependence, the highly anisotropic behavior, the special symmetry, and the 3D dimensional
band overlap. Among the well-stacked graphites, the simple hexagonal (rhombohedral) graphite possesses
the higher (lower) geometric symmetry, the largest (smallest) energy width along ${\hat k_z}$, the
strongest (weakest) band overlap, and the heaviest (lightest) free electron and hole densities. The
unusual geometric and electronic properties are responsible for the unique electronic excitations. As to
the AA-stacked graphite, both low-frequency e-h excitations and plasmon modes could survive for any
directions of transferred momenta, in which the optical modes are similar to those in a 3D electron gas.
The significant differences between the parallel and perpendicular transferred momenta cover the higher
plasmon frequency, the lower peak intensity, and the smaller critical momentum under the former case. It
is relatively easy to observe the low-frequency optical plasmons in the perpendicular case because of the
almost vanishing inter-$\pi$-band e-h excitations. On the other hand, both AB- and ABC-stacked graphite
display the low-frequency optical plasmons only for the perpendicular transferred momenta, in which the
dispersion relations with ${q_z}$  are weak. Their low-frequency excitations are strongly affected by
temperature because of very low free carrier densities, especially for the existence of the optical
plasmons in the latter. The predicted $T$-dependent plasmon frequencies in Bernal graphite are roughly
consistent with those measured by the high-resolution EELS
\cite{ZPhys243;229,PR138;A197,PR178;1340,PRB7;2275,PRL89;076402,OptCom1;119,
SurSci454;462,PRB38;2112} and optical reflectance
spectroscopy.\cite{NanoRes10;234,PR138;A197,PR178;1340,PRB7;2275} Doping in the AA-stacked graphite leads
to drastic changes in the $\pi$-electronic excitations, such as, the great enhancement of plasmon
frequency and strength. The theoretical predictions could account for the EELS and optical measurements
on the optical plasmons in LiC$_6$.\cite{APL77;238,PRB31;4773} The simple hexagonal graphite sharply
contrasts with a monolayer graphene in terms of the low- and middle-frequency excitation channels. For
example, concerning graphene system, the low-frequency plasmon modes are absent for the pristine case and
${T=0}$. They are created by the doping or finite temperatures, but their behaviors at long wave length
limit belong to the acoustic modes. Moreover, the $\pi$-plasmon frequency of graphene is much lower than
those of the 3D systems at small transferred momenta, owing to the lower and narrower saddle-point DOS,
and the weaker Coulomb interactions at small momenta. Specifically, the diversified $\pi$ plasmons are
revealed in three kinds of graphites, in which the main features, the existence, intensity and frequency,
depend on the direction and magnitude of the 3D transferred momenta, and the stacking configurations. The
$\pi$ plasmons, the collective oscillation mode of the whole valence $\pi$ electrons, are present under
any parallel momenta, while those for the perpendicular momenta are present only  at the sufficiently
large parallel components. The dependences on ${q_{\parallel}}$ and ${q_z}$, respectively, grow rapidly
and decline slowly in the increment of momentum. The azimuthal anisotropy comes to exist when
${q_{\parallel}}$ is enough high. The predicted  $\pi$-plasmon frequencies for the
${q_{\parallel}}$-component approximately agree with the EELS measured results on Bernal
graphite.\cite{ZPhys243;229,PR138;A197,PR178;1340,
PRB7;2275,PRL89;076402,OptCom1;119,SurSci454;462,PRB38;2112} In addition, a simple superlattice model is
also utilized to explain the significant features of the $\pi$-electronic collective excitations.
Apparently, such $\pi$ plasmons are revealed in the ${sp^2}$-bonding carbon-based materials, but
regardless of the ${sp}$ and ${sp^3}$ bondings. There exist certain important differences among
graphites,\cite{PLA352;446,OptCom1;119,ZPhys243;229} layered graphenes,
\cite{PRB74;085406,PRB88;075433,PRB98;041408} carbon nanotubes,
\cite{JPSJ66;757,PRB50;17744,PRB47;6617,PRB76;115422,PRB95;195411,Carbon37;733,PRB64;115424,PRL80;4729}
and C$_{60}$-related fullerenes\cite{PRB59;5832}, such as, the momentum- and angular-momentum-decoupled
collective oscillation modes.

The 1D carbon nanotubes have the rich electronic properties and many-particle excitation phenomena, being
sensitive the changes in  the transferred momenta and angular moments (${(q, L)}$),
nanotube geometry (radii $\&$ chiral angles), temperatures, doping levels, and magnitudes $\&$ directions
of the magnetic fields. There are three types of ${(m,n)}$ carbon nanotubes under the curvature effects
according to the concise rule: (1) type-I metals for ${m=n}$ (armchair), (2) type-II narrow-gap systems
for ${m\neq\,n}$ $\&$ ${2m+n=3I}$, and (3) type-III moderate-gap ones for ${2m+n\neq\,3I}$. Only the
first type exhibits the low-frequency acoustic plasmon with the specific ${q|ln(qr)|^{1/2}}$ dependence
at small momenta. Such plasmon comes from the interband excitations of a pair of linearly intersecting
valence and conduction bands near the Fermi level. This plasmon mode hardly depends on finite
temperatures, while the sufficiently high $T$’s can create the intraband low-frequency plasmon in a
narrow-gap large carbon nanotube and a zero-gap semiconducting monolayer graphene. The free carriers can
induce a low-frequency intraband plasmon of ${L=0}$ in all the carbon nanotubes. Furthermore, the
high-level doping even leads to the creation of the ${L=1}$ optical plasmon mode. As to the magnetic
field, its strong effects are to drastically/dramatically alter the energy dispersions $\&$ band gaps of
electronic structures and the wavefunctions, mainly arising from the obvious variation of
${J\rightarrow\,J+\phi\,/\phi_0}$ and the coupling of distinct $J$$^{,}$s. The former mechanism induces
the periodical Aharnov-Bohm effect under the negligible Zeeman splittings, and the latter one results in
the irregular standing waves on a cylindrical surface. Each armchair nanotube only shows one
magnetoplasmon mode due to the interband excitations under the non-perpendicular magnetic field. However,
the low-frequency magnetoplasmon becomes the composite intraband and interband mode at
${\alpha\,=90^\circ}$ or high $T$$^{,}$s. The nonarmchair narrow-gap .nanotubes exhibit two
magnetoplasmon, while the nonarmchair metallic nanotubes present one interband magnetoplasmon and one
interband $\&$ intraband magnetoplasmom. The Zeeman splitting plays an important role on the significant
differences among these magnetoplasmon modes. In addition to the low-frequency $\pi$-electronic
excitations, there exist the inter-$\pi$ band and $\pi$ plasmons in various carbon nanotubes, in which
they, respectively, corresponds to the specific valence bands and the whole $\pi$-valence electrons.
Their dependence on angular momentum, radius and chiral angle is significant, while the momentum
dispersion relation might be strong  or weak. Among three types of nanotube systems, armchair/zigzag ones
have the smallest/largest mode number of inter-$\pi$-band plasmons. The $\pi$-plasmon frequencies of
${\omega_p\,>5}$ eV, which grow with the increasing ${(q, L)}$, are very suitable in identifying the
intrinsic $L$-decoupled modes of the cylindrical systems. Up to now, the higher-frequency
inter-$\pi$-band and $\pi$ plamsons have been confirmed by the accurate EELS
measurements.\cite{PRB95;195411,Carbon37;733,PRB64;195404,
PRB64;115424,PRL80;4729,JJAP31;L1484,PRB49;2882} However, the low-frequency ones, being due to the
geometry, temperature, magnetic field and doping, require the further experimental verifications.
Apparently, the Coulomb excitation phenomena are greatly diversified by the 1D cylindrical surface and
the 2D plane (the different geometric symmetries). The distinct bare Coulomb interactions and electronic
properties can create the sharp differences for carbon nanotubes and layered graphenes, covering the
conserved quantities in the electron-electron scatterings, the existence/absence of the low-frequency
plasmons, and the composite effects coming from the critical factors (temperature, strength and direction
of magnetic field, and doping density).

Obviously, group-IV monolayer systems exhibits the rich and unique Coulomb excitation spectra, in which
there are certain important differences among silicene, germanene and graphene. For pristine systems, the
near-neighboring hopping integral, spin-orbital coupling and temperature play the critical roles in
creating the available single-particle transitions and the observable plasmon modes. The first and second
factors are, respectively, responsible for and gap and Fermi velocity; therefore, they co-determine the
critical temperature and the threshold excitation frequency. When the temperature is sufficiently high,
the $T$-induced intraband transitions will strongly suppress and even replace the original interband
ones. The intraband and interband e-h excitations, respectively, display the square-root divergent peak
and shoulder structure in the imaginary part of dielectric function. There are three types of collective
excitations, as clearly illustrated from ${(q, \omega\,)}$ phase diagrams under the specific
temperatures. For example, silicene, with a very narrow gap, shows (I) the undamped plasmon between the
intraband and interband boundaries, (II) the similar one and another weaker plasmon and (III) the
undamped and damped plasmon, in which the first and third kinds of plasmon modes (the second kind)
experiences the interband (intraband) Landau dampings. The diverse behaviors strongly depend on the range
of temperature: (I) ${T<}$75 K, (II) 75 K${\le\,T\le\,120}$ K and (III)  ${T>120}$ K. The electric field
in a buckled system can destroy the mirror symmetry about the ${z=0}$ plane and create the splitting of
spin-up and spin-down energy bands, resulting in the diversified Coulomb excitation phenomena in the
${(q, \omega\,)}$-, ${(T, \omega\,)}$- and ${(F, \omega\,)}$-phase diagrams. They clearly present the
different critical momenta, temperature and electric-field strengths under the various cases, indicating
the dramatic changes of the intensities of energy loss functions and thus the emergence/disappearance of
plasmon mode and the significant Landau dampings. Furthermore, four types of single-particle and
collective excitations, which depend on the specific $T$ and $F$, are characterized by: (I) the
excitation gap exists between the intraband and interband transitions of the spin-degenerate valence and
conduction bands (without $F$), and an intraband plasmon is undamped at small $q$$^{,}$s and then
disappears at large $q$$^{,}$s because of the serious Landau dampins near the intraband boundary. (II)
three e-h boundaries, the lower, middle and upper ones, are induced by the intraband transitions from the
spin-down-dominated energy bands, the interband transitions of the same pair, and the interband
transitions due to the spin-up-dominated pair. The most prominent intraband plasmon survives between the
former two boundaries. It experiences the interband e-h damping near the  middle boundary, accompanies
with another weaker plasmon. This mode is absent at large $q$$^{,}$s under the intraband dampings (close
to the lower boundary). (III) the merged intraband and interband boundaries related to the linear energy
bands intersecting at ${E_F}$ and the higher-frequency interband boundary associated  with the
spin-up-dominated energy bands. The intraband plasmon has a strong e-h dampings even at long wavelength
limit. (IV) The e-h boundaries are the same with those in (II), while this plasmon is undamped only near
${q\rightarrow\,0}$, and then it enters into the distinct interband Landau dampings related to the first
and second pairs of spin-split energy bands.
Concerning the complex composite effects arising from the spin-orbital interactions, magnetic field,
electric field, and electron doping, they further diversify the single-particle and collective
excitations. The magnetic field creates the interband magnetoplasmons with discrete frequency dispersions
restricted to the quantized LL states. An intraband magneoplasmon, with a higher intensity and continuous
dispersion relation, comes to exist in the presence of free conduction carriers. This mode is
dramatically transformed into an interband plasma excitation when the magnetic field is increased,
leading to abrupt changes in plasma frequency and intensity. Specifically, an electric field could
separate the spin and valley polarizations and induce additional magnetoplasmon modes, an unique feature
arising from the buckled structure and the existence of the significant spin-orbit couplings. The
intraband and interband magnetoplasmons, respectively, belong to the continuous and localized modes. In
short, it is relatively easy to observe the split valleys and spin configurations in monolayer germanene,
since it has the largest spin-orbital couplings among three group-IV systems (C, Si; Ge).

Apparently, monolayer graphene presents the rich and unique Coulomb deexcitation phenomena, being very
different from 2D electron gas, 1D carbon nanotubes, silicene, and germanene. The inelastic Coulomb
scatterings could be investigated from the screened exchange self-energy method and the Fermi golden
rule. Concerning the Coulomb decay rates in pristine graphene, they directly reflect the characteristics
of band structure, the zero-gap semiconductor and the strong dependence on wave vector/state energy. At
$T$ = 0, only the interband e-h excitations exist, and their contributions to the Coulomb decay rates are
negligible. The temperature-induced free carriers create the intraband e-h excitations and the acoustic
plasmon. The strength of plasmon mode grows (declines) as temperature (momentum) increases. The Coulomb
inelastic scatterings of the Fermi-momentum state (the Dirac point) only utilize the intraband e-h
excitations, with any transferred momenta, leading to the non-specific $T$-dependence. The calculated
decay rate is close to the measured results of the layered graphite.\cite{PLA357;401,PRB87;085406} The
intraband e-h excitations also make important contributions to other states. Both interband e-h
excitations and plasmon belong to the critical deexcitation channels for the small-$k$ states. They,
respectively, result in the special shoulder and peak structures in the wave-vector-dependent decay rates
of the conduction- and valence-band states. The e-e interactions are more efficient in carrier
deexcitations, compared with the electron-phonon interactions. The important differences between graphene
and 2D electron gas lie in the temperature and wave-vector dependences, since the the band structure of
the latter can only generate the intraband e-h excitations and its acoustic plasmon survives even at
$T=0$. The plasmon is not the effective deexcitation channel and the small-${(q,\omega_{de})}$ intraband
e-h excitations predominate over the relaxation process, so that the temperature- and energy-dependent
Coulomb decay rates could be expressed in the analytic formulas.\cite{PLA357;401} The absence of the
specific formulas in  monolayer graphene clearly indicates that the large-($q,\omega_{de}$) intraband e-h
excitations, the plasmon, and the interband e-h excitations are not negligible in the inelastic Coulomb
scatterings. These three kinds of electronic excitations are expected to play an important role on the
low-energy quasiparticle properties of few-layer graphene systems. There also exist significant
differences between 2D graphene and 1D gapless carbon nanotubes, covering the magnitude,
temperature-dependence and wave-vector-dependence of the Coulomb decay rates. The Dirac point in armchair
nanotube shows the linear $T$-dependence, as observed in a 1D electron gas.\cite{Mahan} Apparently, the
Coulomb decay rates of monolayer graphene are greatly diversified by the electron (or hole) doping. The
deexcitation processes cover the intraband single-particle excitations, the interband e-h excitations,
and the damped or undamped collective excitations. depending on the quasiparticle states and the Fermi
energies. The low-lying valence holes can decay through the undamped collective excitations; therefore,
they present the fast Coulomb deexcitations, nonmonotonous energy dependence, and anisotropic behavior.
However, the low-energy conduction electrons and holes are similar to those in a two-dimensional electron
gas. In addition, the Fermi-momentum states display the ${T^2ln[T]}$ dependence. The higher-energy
conduction states and the deeper-energy valence ones behave similarly in the available deexcitation
channels and have a strong dependence on the wave vector ${\bf k}$, since the contributions due to
intraband (interband) e-h excitations quite differ from one another along the distinct directions.
Moreover, there exist the significant differences among the extrinsic graphene, silicene and germanene in
terms of the available deexcitation channels and certain Coulomb decay phenomena. Specially, a doped
germanene shows the composite effects due to the spin-orbital couplings and electron doping. The second
and third kinds of plasmon modes might be the effective Coulomb scattering channels. Moreover, the
deexcitation mechanisms sharply contrast for the separated valence and conduction Dirac points, in which
the former could decay by the acoustic plasmon and thus has the faster Coulomb decay rate.

The generalized tight-binding model and the modified RPA are developed to deal with the Coulomb
excitations and decay rates under various stacking configurations, layer numbers, dimensionalities,
electric fields and magnetic fields. The bare and screened response functions are thoroughly clarified by
using the dynamically electron-electron inelastic scatterings. Specifically, the latter is very useful in
the further understanding of the time-dependent carrier oscillation/propagation in the real
graphene-related systems. These energy loss spectra are very sensitive to the magnitude and direction of
the transferred momenta under various frequencies. From the theoretical point of view, one must preform
the 3D Fourier transform for them to explore the time- and position-dependent collective oscillation
phenomena accompanied with the Landau dampings. The calculated results in Chaps. 3-14 clearly show that
there are three kinds of plasmon modes due to the carbon ${2p_z}$ orbitals in graphene-related
${sp^2}$-bonding systems. The $\pi$ and inter-$\pi$-band plasmons possess the oscillational frequencies
more than 1 eV, corresponding to a very rapid oscillation/propagation with a period below 4 fs.
Apparently, the time resolution is too high for the up-to-date femtosecond optical spectroscopies; that
is, it is impossible to observe and identify the oscillation behaviors from these two kinds of plasmon
modes using the current high-resolution equipments. On the other hand, the low-frequency plasmon modes,
with the different momentum dispersions, are expected to reveal in all the graphene-related materials.
For example, the ${\sim\,10}$-meV plasmon modes, with oscillation periods ${\sim\,400}$ fs, are very
suitable for the experimental verifications on the diverse plasmon waves accompanied with the various
Landau dampings. Such collective excitations are deduced to originate from the temperatures, lattice
symmetry, dimensionalities, stacking configurations, numbers of layers, spin-orbital interactions,
electric fields, and magnetic fields. That is to say, a lot of critical mechanisms could also be
illustrated by the dynamic behavior of the excited carriers.
The time-dependent charge oscillations/propagations in the layered structures will become an interesting
and emergent research topic; furthermore, the dynamics of plasmon modes might have high potentials in
electronic and optical devices.\cite{NanoLett13;5991,NatPhys3;36} Such work is under current
investigation.

The above-mentioned two models are very useful in fully exploring the static charge screening of
graphene-related systems, especially for the spatial variations (distributions) of the effective Coulomb
potentials and the induced carrier density on the distinct planes. The momentum- and frequency-dependent
Coulomb potentials, which have been thoroughly investigated in the current work for the 1D, 2D and 3D
${sp^2}$-bonding systems, could be recovered to the static effective Coulomb  interactions in the real
space by doing the Fourier transform at zero frequency. The similar method is applied to the induced
charge density (the product of the effective potential and the bare response function in the momentum
space). The free carrier density and the Fermi surface (the Fermi momenta), or the energy gap are
expected to play critical roles on the unusual screening behaviors. Whether the long-range effective
Coulomb interactions decay quicker than the inverse of distance is the standard criteria for the charge
screening ability.
The semiconducting carbon nanotubes have been predicted to exhibit the long-distance Coulomb potential
similar to the bare one.\cite{PRB56;4996} However, the metallic armchair nanotubes\cite{PRB76;115422},
layered graphenes\cite{PRB82;193405} and graphite intercalation compounds\cite{PRB46;12656} obviously
display the exponential decay at the short distance and the well-known Friedel oscillations at the long
distance. The characteristic decay length and the rapid charge oscillations are very sensitive to the
concentration of free carriers and the spatial dimensions. From the theoretical point of view, the strong
Friedel oscillations mainly come from the divergent derivative of the static bare response functions
versus the transferred momentum at the double of Fermi momentum, as identified from 3D and 2D electron
gases.\cite{Mahan}
On the other hand, STM has been demonstrated to be a powerful tool for visualizing the Friedel
oscillations in real space of 2D graphene-related systems in response to the atomic-scale
impurities.\cite{RMP81;1495}
In general, both pristine few-layer graphene systems and 3D graphites, which belong to the semi-metals,
are expected to show the marginal (partial) screening ability in between the full and almost vanishing
ones. Also, the doped layered graphene systems, with the distinct stacking configurations and layer
numbers, are very suitable for exploring the diversified charge screening phenomena, e.g., the
theoretical\cite{PRB82;195428,PRB75;205418,PRB84;115420,PRB93;035440,
PRB86;195424,JPAMT42;214015,PRB84;085112,PRB74;085406,PRB56;4996} and experimental studies on the 2D
Friedel charge oscillations.\cite{PRL98;206802,PRL110;146802,PRB89;195410} In addition, the effective
Coulomb potentials due to the charged impurities could investigate the residual resistivities in layered
materials, e.g., the zero-temperature elastic scatterings for the residual resitivies in alkali-doped
graphite compounds.\cite{PRB46;12656}

The Friedel oscillations can be studied with the use of the static dielectric screening in the
generalized tight-binding model. In a monolayer graphene, the static dielectric function $\epsilon(q,0)$
is calculated by specifying the general results of the full charge screening due to the massless-Dirac
quasiparticle, which is incorporated by the development of the modified random-phase approximation (RPA),
summarized in Chap. 2.
As shown in Fig. 14-11, the induced charge distribution and the screened potential are presented for the
static case of a charge impurity $e$ at $r=0$ in the doped graphene under various Fermi levels.
The most outstanding feature of the Friedel oscillations is the strong oscillation behaviour in real
space, a phenomenon being deduce from a discontinuity of the second derivative of the static dielectric
function.
At higher dopings, the screening charges become rather extended in r-space, with a $1/r$ long-wavelength
decay of oscillations at a distance from the impurity.\cite{PRB82;193405}
It should be noticed that there is no interband contribution from the long-wavelength behaviour of the
polarization since the corresponding polarization approaches zero at $q\sim 0$.
This indicates that the non-oscillating part at small-r distances comes from the intraband polarization
with a characteristic decaying length that can be evaluated within the Thomas-Fermi approximation.
The previous work has indicated that as $q$ is increased the interband transitions greatly enhance the
large-$q$ screening as compared to the intraband transitions, and consequently make the small-$r$
screening effects rather effective from an impurity.\cite{PRB46;12656}
However, and very importantly, at a long distance from the impurity, the decaying oscillation behaviour
with a unique period given by $\pi/2k_{F}$ comes from the discontinuity at $\hbar v_{F} q=2 E_{F}$ as a
result of the singularities in the second derivative of the static dielectric function.
The aforementioned features are expected to have important influences on, for example, the ordering of
impurities and the resistivity in graphene-related systems.\cite{PRB46;12656}

The dimensionless energy loss function, which characterizes the intrinsic charge screeing of a layered
condensed-matter system, is well developed in this work. Furthermore, the modified RPA [Chap. 2.1] is
very suitable and reliable in evaluating the effective Coulomb potentials for the intralayer and
interlayer many-particle interactions. As a result, such potentials could be directly linked with the
self-energy method of the Matibara Green functions [Chap. 2.6]. The decay rate, being associated with the
e-e Coulomb inelastic scatterings on all the graphene layers, could be derived and even expressed in the
analytic form. The phenomenological formulas will be investigated fully and delicately. The effects
arising from the number of layers and stacking configurations are taken into consideration
simultaneously. The rich and unique deexcitation phenomena are expected to reveal in few-layer graphene
systems, and they need to be thoroughly explored  within this new theoretical framework. For example, the
AAA-/ABA-/ABC-stacked $N$-layer graphenes are present in an electromagnetic field, the excited conduction
electrons and valence holes in $N$ pairs of energy bands will relax through the effective Coulomb
excitation channels. The numerical calculations become very difficult, since the anisotropic
muti-dimensional integrations appear in the band structure, polarization functions, and transferred
momenta and energies. Some efficient methods need to be introduced to overcome the serious issues. On the
experimental side, the femtosecond pump-probe optical spectroscopies of absorption and fluorescence are
available to identify the diverse deexcition phenomena in few-layer graphene systems. Apparently, such
studies could be generalized to the other emergent 2D materials. In short, the near-future researches on
the femtosecond dynamics of charge carriers are predicted to open a new studying category.

In addition to the Coulomb scatterings, the electron-phonon scatterings at finite temperatures might play
important roles in the electron-electron effective interactions and thus have a strong effect electronic
excitation spectra and quasiparticle lifetimes. These two kinds of inelastic scatterings could be taken
into account simultaneously under the random-phase approximation of the same order, as clearly
illustrated in Refs.\cite{Mahan}.
This method has been clearly verified to be suitable and reliable for the doped GaAs semiconductor by
measuring the plasmon branches using the Raman scattering.\cite{PRL16;999}
Apparently, the combination of the longitudinal plasmon modes and optical phonons creates the well-known
anti-crossing phenomenon in the spectrum of the collective oscillation frequencies.
In the near-future studies on few-layer graphene systems, the couplings of acoustic plasmons and
longitudinal optical phonons are expected to diversify the (momentum, frequency)-phase diagraoms and
greatly modify the inelastic scattering rates of the excited quasiparticle
states.\cite{NanoLett17;5908,ACSPhotonics4;2908,PRB81;081406,Science328;999}
On the other side, certain physical barriers in theoretical models need to be overcome, such as,
the delicate phonon spectra of layered graphenes calculated by the oscillator
model,\cite{PRB86;165422,PRB75;155420}
the effective amplitude of the electron-phonon scatterings using the phenomenological
method,\cite{JPSJ76;104711,PhysE40;213,PRB45;768}
the modifications of the layer-dependent random-phase approximation,\cite{PRB42;195406}
and the modified screened exchange energies.\cite{PRB77;081411}
They will become challenges and chances for a full understanding of the overall quasipaticle phenomena in
2D emergent materials. For example, there are some theoretical predictions on phonon spectra, especially
for the first-principles calculations,\cite{PRB77;125401,NanoLett8;4229,PRB81;121412} in which how to
transform the evaluated results into the very useful information in establishing the electron-phonon
scattering amplitudes is one of the  high barriers.

\par\noindent {\bf Acknowledgments}

This work was supported in part by the National Science Council of Taiwan,
the Republic of China, under Grant Nos. NSC 98-2112-M-006-013-MY4 and NSC 99-2112-M-165-001-MY3.

\newpage
\renewcommand{\baselinestretch}{0.2}

\end{document}